\documentclass[%
 amsmath,amssymb,
 aps,
 prb,
 twocolumn,
 superscriptaddress,
 longbibliography
]{revtex4-2}

\usepackage{graphicx}
\usepackage{dcolumn}
\usepackage{bm}
\usepackage{xcolor}
\usepackage{tikz}
\usepackage{xfrac}
\usepackage{blindtext}
\usepackage{times}
\usepackage{multirow}
\usepackage{chemformula}
\usepackage{braket}
\usepackage{tabularx}

\usepackage{color}

\usepackage[normalem]{ulem}

\definecolor{darkblue}{HTML}{004D6B}
\definecolor{darkred}{HTML}{8c1515}
\definecolor{darkgreen}{HTML}{006400}
\usepackage{hyperref}
\hypersetup{
	colorlinks=true,
	urlcolor=darkred,
	citecolor=darkblue,
	linkcolor=darkred,
	breaklinks
}

\usepackage[caption=false]{subfig}

\graphicspath{{figures/}}

\newcommand{\be}{\begin{equation}}
\newcommand{\ee}{\end{equation}}

\newcommand{\bea}{\begin{eqnarray}}
\newcommand{\eea}{\end{eqnarray}}
\newcommand{\beal}{\begin{align}}
\newcommand{\eeal}{\end{align}}

\begin{document}

\title{Homological Thresholds in Randomly Monitored  Quantum Error-Correcting Codes}

\author{Yoshito Watanabe}
\affiliation{Institute for Theoretical Physics, University of Cologne, 50937 Cologne, Germany}

\author{Daiki Sasamoto}
\affiliation{Department of Physics, Graduate School of Science, Tohoku University, Sendai, Miyagi 980-8578, Japan}
\affiliation{Institute for Theoretical Physics, University of Cologne, 50937 Cologne, Germany}

\author{Bo Han}
\affiliation{Institute for Theoretical Physics, University of Cologne, 50937 Cologne, Germany}

\author{Simon Trebst}
\affiliation{Institute for Theoretical Physics, University of Cologne, 50937 Cologne, Germany}

\date{\today}

\begin{abstract}
Random projective measurements are an effective tool to model the interlinked phenomenology of monitoring, inferring, and decoding highly entangled states such as quantum error-correcting codes.
At the same time, the study of such monitored dynamics has opened a new arena for critical phenomena, 
renewing longstanding questions in disordered statistical mechanics, including percolation, random-bond models, and Nishimori physics. 
Here we investigate the robustness of stabilizer quantum error-correcting codes under a single round of independently sampled single-qubit Pauli measurements
and demonstrate that their threshold behavior falls into two broad categories -- {\sl geometric} and {\sl homological} percolation transitions, with the toric code 
and color code being the principal examples for each class. 
We establish the {\sl distinct} character of these two threshold theories by analyzing the logical-support criteria of these codes 
and by providing numerical estimates of their respective critical exponents obtained from optimized stabilizer simulations of large-scale systems with over one million qubits.
Our framework extends to non-CSS, subsystem, higher-dimensional, and Floquet stabilizer codes
and allows us to determine their respective threshold phase diagrams. 
We exemplify how the latter provide quantitative and qualitative guidance for the computationally more demanding problems 
of tunable learning and maximum-likelihood decoding.
As a practical application of our framework, we discuss the robustness of the gross code.
\end{abstract}

\maketitle

\section{Introduction}

Stabilizer codes \cite{Gottesman1996,Gottesman1997} are a foundational principle of quantum information theory, 
establishing how one can create and actively protect a logical subspace 
via a set of multiqubit stabilizer measurements that feed into a quantum error correction scheme \cite{Nielsen2010}. 
Stabilizer codes are also of conceptual interest in quantum many-body physics
where they provide an efficient path to prepare a macroscopically entangled quantum state \cite{Kitaev2003,Briegel2001,Raussendorf2005,Verresen2022,Tantivasadakarn2024}
whose robustness can subsequently be probed in the presence of decoherence, inference,
or monitoring. 
One early example linking both perspectives is the determination of an error threshold
for the toric code in the presence of incoherent noise \cite{Dennis2002}. 
Via a joint connection  to the statistical mechanics of the random-bond Ising model \cite{Chen2025IBM}
and its Nishimori line \cite{Nishimori1981}, this decoherence transition is closely linked to an
inference transition \cite{Eckstein2024,Eckstein2025}, which is induced by weak single-qubit 
measurements aiming to learn the logical subspace of the toric code.
Finally, monitoring the toric code by a round of random single-qubit measurements (that do
not commute with the stabilizer measurements) can also collapse the logical subspace 
at finite read-out probability~\cite{Negari2024,Botzung2025,Lee2026}, in a transition captured by percolation theory.

\begin{figure*}
    \centering
    \includegraphics[width=1.0\textwidth]{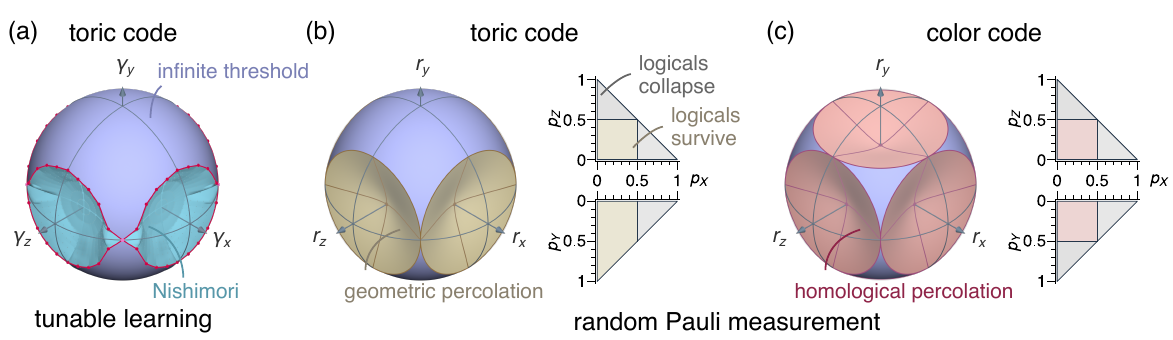}
    \caption{
    {\bf Tunable-learning and random-measurement phase diagrams.}
    (a) Toric-code tunable-learning phase diagram of Ref.~\cite{Eckstein2025}, parameterized by the measurement strengths $(\gamma_X,\gamma_Y,\gamma_Z)$. The cyan boundary marks the Nishimori learning transition. (b,c) Semi-analytic phase diagrams for the random Pauli-measurement protocol studied here, for (b) the toric code and (c) the color code. The phase boundaries separate regimes in which encoded logical information survives from those in which it is collapsed by the measurements. The measurement probabilities are parameterized as $p_X=r\sin^2\theta\sin^2\phi$, $p_Y=r\cos^2\theta$, and $p_Z=r\sin^2\theta\cos^2\phi$, with $r=p_X+p_Y+p_Z$. The boundaries are approximate constructions based on analytically obtained pure-axis thresholds. Purple marks the radial endpoint: an infinite learning threshold in (a) and the fully measured limit $r=1$ in (b,c). Yellow and red denote geometric and non-geometric homological percolation, respectively. The triangular insets show the same approximate boundaries directly on the $p_Y=0$ $(p_X,p_Z)$ and $p_Z=0$ $(p_X,p_Y)$ planes; colored and gray regions indicate logical survival and collapse, respectively.
    }
    \label{fig:phase_diagram}
\end{figure*}

In this manuscript, we discuss how the robustness of a stabilizer code in these three different 
scenarios is interconnected and what kind of quantitative or qualitative  guidance one has 
going from one setting to another -- for instance, how an easy-to-calculate percolation threshold
for monitoring allows one to bound the Nishimori thresholds for weak measurements and to 
readily deduce the principal shape of a complex and hard-to-calculate inference phase diagram, 
cf.\ Figs.~\ref{fig:phase_diagram} and \ref{fig:overlay}. 
Or how the decoding threshold in the presence of incoherent noise can be bounded by both 
the monitoring and inference thresholds.

We start by mapping out the robustness of a broad family of stabilizer codes in the presence
of random monitoring where each physical qubit is independently measured in the $X$, $Y$, or $Z$ basis 
with probabilities $p_X$, $p_Y$, and $p_Z$, respectively, or is left unmeasured with probability $1-p_X-p_Y-p_Z$,
giving rise to spherical phase diagrams such as the ones shown in Fig.~\ref{fig:family} that separate
a regime in which the logical subspace remains intact from regimes in which the measurement record 
fixes one or more nontrivial logical observables.

For this monitoring setup, the underlying organizing principle is that the measured Pauli operators generate a commuting group containing many different products. Some act trivially on the encoded degrees of freedom, while others may implement logical Pauli measurements.  Logical information is lost when this measurement-generated group contains a nontrivial logical Pauli operator, up to multiplication by stabilizers. This criterion is determined by {\sl logical support} rather than {\sl geometric connectivity} and marks the difference between a threshold theory that is captured by {\sl homological} versus ordinary {\sl geometric} percolation.

The toric and color codes illustrate this distinction, cf.~Fig.~\ref{fig:schematics}. In the pure-$Z$ limit, both exhibit a transition at $p_Z=1/2$. For the toric code, the problem reduces to ordinary bond percolation on the square lattice~\cite{Kesten1980}, and the extracted critical exponent $1.334(43)$ is consistent with the two-dimensional percolation universality class. For the color code, the threshold is fixed instead by the $X/Z$ symmetry and complementary-set structure of the code. The associated logical-support problem is non-geometric, and finite-size scaling gives $\nu=1.2073(15)$, suggesting critical behavior distinct from ordinary percolation. Thus, identical threshold locations need not imply identical critical behavior. 

The analytically determined pure-axis thresholds also provide a simple approximation to the mixed-measurement phase boundaries shown in Fig.~\ref{fig:phase_diagram}, 
allowing one to basically map out the entire spherical phase diagram. Indeed, our numerical results attest to the validity of such an approximation [Fig.~\ref{fig:numerical_data}], 
while also indicating subtle limitations, such as cooperative effects involving different Pauli measurements, cf.~Fig.~\ref{fig:y_assisted_logical_z_measurement}.

Taking a step back, our results establish homological percolation as a unifying language for logical-survival transitions across a broad range of randomly monitored stabilizer codes, including non-CSS, higher-dimensional, subsystem, and Floquet settings, while revealing diverse critical behavior within this common framework. 
The homological organization also extends to the setups of inference/tunable learning and maximum-likelihood QEC decoding, as we will discuss in Sec.~\ref{sec:wider_perspective}. 
It makes the relatively easy-to-handle projective-monitoring problem a tractable window onto a broader landscape, where resolving the phase structure entails substantially more demanding partition-function calculations.

\subsection*{Random-Measurement Protocol}

We define the protocol as follows. Each physical qubit undergoes a single round of random single-qubit Pauli measurements, with probabilities parameterized by $(\theta,\phi,r)$ as
\begin{equation}
  \begin{split}
    &p_X=r\sin^2\theta\sin^2\phi,\quad p_Y=r\cos^2\theta,\\
    &p_Z=r\sin^2\theta\cos^2\phi,\quad p_I=1-r.
  \end{split}
\end{equation}
Here $r=p_X+p_Y+p_Z$ is the total measurement probability, while $p_I$ is the probability that a qubit is left unmeasured. For the spherical phase diagrams, we introduce coordinates
\begin{equation}
  \begin{split}
    &r_X=r\sin\theta\sin\phi,\quad r_Y=r\cos\theta,\\
    &r_Z=r\sin\theta\cos\phi.
  \end{split}
  \label{eq:spherical_coordinates}
\end{equation}
Thus $r=\sqrt{r_X^2+r_Y^2+r_Z^2}$, and the measurement probabilities satisfy $p_\alpha=r_\alpha^2/r$ for $r>0$, with $\alpha\in\{X,Y,Z\}$. Along a pure-$\alpha$ axis, $p_\alpha=r_\alpha=r$. We then ask whether the encoded logical information survives this single round of random measurements.

To locate the phase boundaries~\cite{Colmenarez2024}, we maximally entangle the $k$ logical qubits of the code $Q$ with $k$ reference qubits $R$ and compute the coherent information~\cite{Schumacher1996,Lloyd1997}. We retain the complete classical measurement record $C$, including measurement locations, Pauli bases, and outcomes. The coherent information~\cite{Colmenarez2024,Schumacher1996,Lloyd1997} is given by
\begin{equation}
\begin{aligned}
I_c(R | QC)
=S(QC)-S(RQC)
=\sum_c p_c S\!\left(\rho_R(c)\right) \,,
\end{aligned}
\label{eq:coherent_information_record}
\end{equation}
where $S$ denotes the von Neumann entropy in bits, $p_c$ is the probability of record $c$, and $\rho_R(c)$ is the corresponding conditional reference state. The second equality follows because the code and reference remain jointly pure for each complete measurement record. For projective Pauli measurements, this reference entropy counts the surviving logical Bell pairs in each realization, so $I_c$ gives their average number.

For a fixed measurement realization, the measurement record also determines the outcome of any product of the measured Pauli operators. Logical information is lost when such a product is equivalent, up to stabilizers, to a nontrivial logical Pauli operator. In that case, the corresponding logical Pauli eigenvalue has been measured, and the encoded logical space is reduced.  
The same condition can be reformulated in the language of CSS chain complexes and symplectic stabilizer complexes, see Appendix~\ref{app:homological_percolation}.

\section{Toric code}

We begin with the toric code~\cite{Kitaev2003}, defined on a square lattice with periodic boundary conditions and qubits placed on the edges. The stabilizers are vertex operators $A_v$ and plaquette operators $B_p$, given by products of $X$ and $Z$ Pauli operators, respectively,
\begin{equation}
  A_v=\prod_{e \in \delta v} X_e,
  \qquad
  B_p=\prod_{e \in \partial p} Z_e,
\end{equation}
where $\delta v$ is the set of edges incident to vertex $v$ (the coboundary of $v$) and $\partial p$ is the boundary of plaquette $p$. On the torus, the code encodes two logical qubits. Logical $Z$ operators are represented by noncontractible loops of $Z$ operators on the direct lattice, while logical $X$ operators are represented by noncontractible loops of $X$ operators on the dual lattice.

\subsubsection*{$Z$-only limit}

\begin{figure}[t]
    \centering
    \includegraphics[width=1.0\columnwidth]{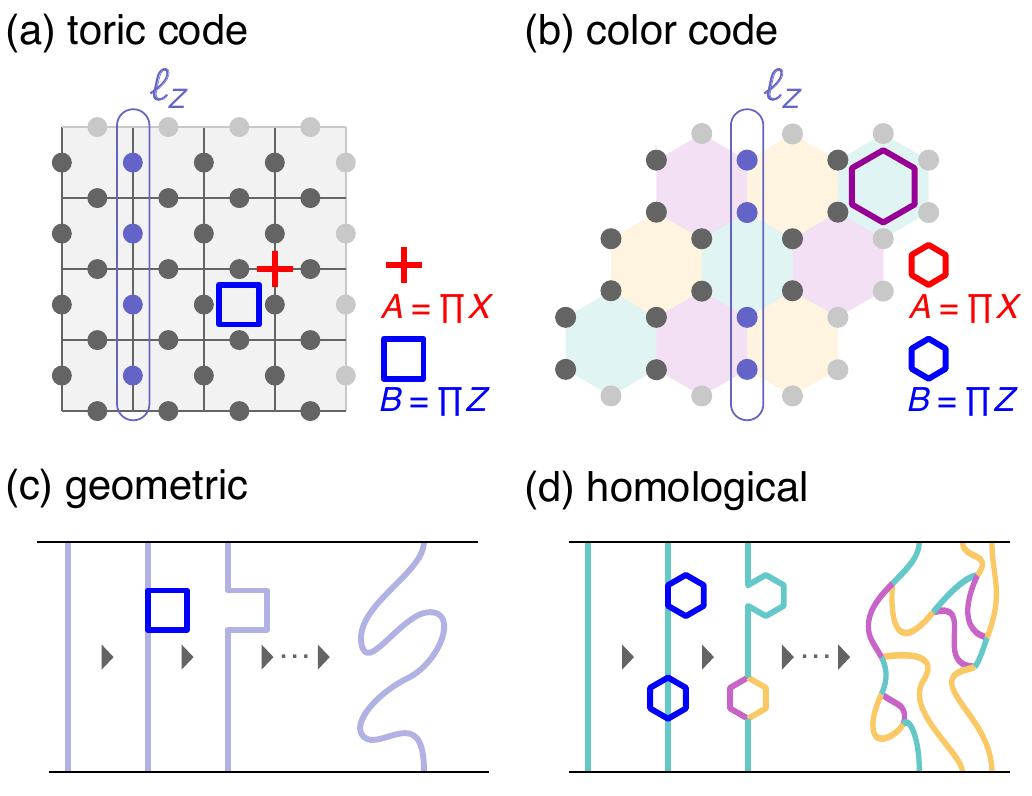}
    \caption{
    {\bf Local structure of logical operators.}
    (a,b) Square-lattice toric code and honeycomb-lattice $(6,6,6)$ color code, with representative noncontractible logical $Z$ operators. Red and blue symbols denote $X$- and $Z$-type stabilizers, respectively.
    (c,d) Multiplication by local $Z$ stabilizers deforms these representatives without changing their logical action. Toric-code deformations preserve an unbranched-loop description, whereas color-code strings can split and recombine at three-string junctions, producing branched colored-string representations. This contrast underlies ordinary geometric connectivity in the toric code and non-geometric homological connectivity in the color code.
    }
    \label{fig:schematics}
\end{figure}

In this limit, the logical subspace collapses only if the measured $Z$ edges support a nontrivial logical $Z$ operator. In the toric code, this means that the measured edges must contain a noncontractible loop, possibly after deforming it by plaquette stabilizers, see Fig.~\ref{fig:schematics}(c). The transition therefore reduces to ordinary bond percolation on the square lattice, with critical point $(p_Z)_{\mathrm c}=1/2$~\cite{Kesten1980}.

With $N$ physical qubits and $F$ plaquettes, we work over $\mathbb F_2$ and define
\begin{equation}
  D\in\mathbb F_2^{N \times F},
  \qquad
  D_{ep}=1
  \quad\Longleftrightarrow\quad
  e\in\partial p \, .
\end{equation}
Each column of $D$ contains the four edges bounding one plaquette, so $D$ is the plaquette-boundary map. A vector $a\in\mathbb F_2^F$ specifies a product of plaquette stabilizers, whose support is $Da$. Let $\ell\in\mathbb F_2^N$ be the support of a reference noncontractible $Z$ loop. Multiplication by the plaquette stabilizers specified by $a$ changes its representative according to
\begin{equation}
  \ell\longmapsto\ell+Da \,.
\end{equation}
This operation is illustrated in Fig.~\ref{fig:schematics}(c). Each local stabilizer multiplication reroutes the loop without introducing a branch or changing its winding class.

The measured edges form $M$, and the unmeasured edges form $U=\overline M$. The logical class represented by $\ell$ is measured exactly when stabilizer multiplications move its entire support onto $M$. This amounts to finding a stabilizer pattern $a$ satisfying
\begin{equation}
  D_Ua=\ell_U ,
  \label{eq:toric_code_du_condition}
\end{equation}
where $D_U$ and $\ell_U$ denote the restrictions to the unmeasured edges. Because the allowed deformations are precisely the local reroutings generated by the physical plaquette boundaries, Eq.~\eqref{eq:toric_code_du_condition} is equivalent to asking whether the occupied bonds contain a noncontractible loop. The toric-code transition is therefore the {\sl geometric realization} of homological percolation. Finite-size scaling within the stabilizer-tableau formalism gives $\nu_\mathrm{stab}=1.334(43)$, consistent with the exact two-dimensional bond-percolation value $\nu=4/3$.

As we will see in Sec.~\ref{sec:color_code}, the same algebraic support criterion leads to a different local deformation rule in the color code, as illustrated in Fig.~\ref{fig:schematics}(d). Consequently, the problem no longer reduces to ordinary path connectivity.

\begin{figure}[b]
    \centering
    \includegraphics[width=1.0\columnwidth]{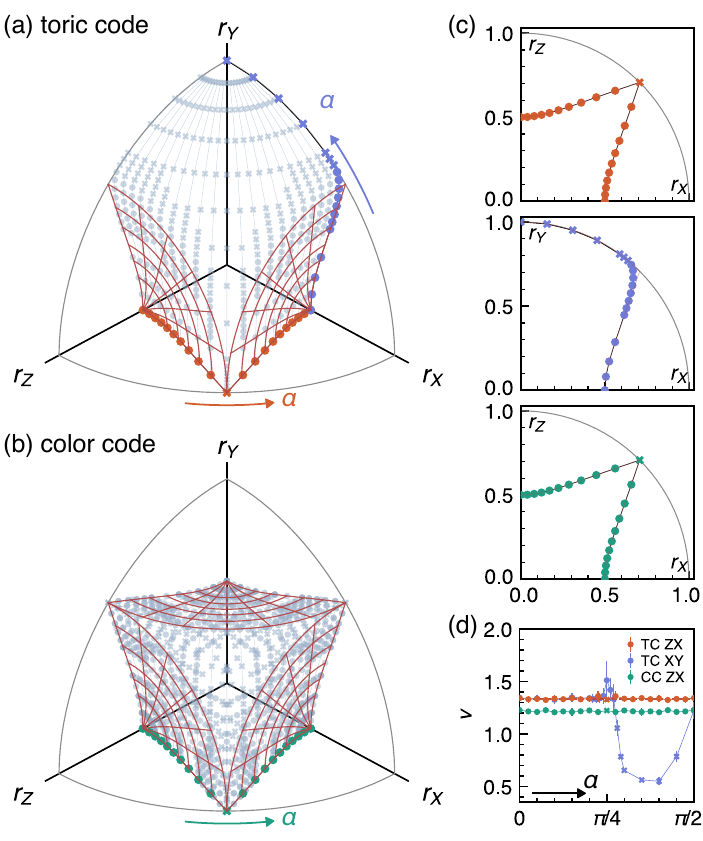}
    \caption{
    {\bf Numerical logical-survival phase diagrams.}
    Phase boundaries for (a) the toric code and (b) the color code are shown in the positive octant. Red wireframes show the semi-analytic construction from the pure-axis thresholds, while colored symbols show numerical boundary estimates obtained from finite-size scaling of the coherent information. Crosses mark estimates compatible with $r_{\mathrm c}=1$ within uncertainty; the corresponding thermodynamic critical radii may lie slightly below unity (see text). (c) Representative planar cuts.
    (d) The fitted scaling exponent $\nu$ along the same cuts as a function of the in-plane angle $\alpha$, oriented as indicated in (a,b).
    }
    \label{fig:numerical_data}
\end{figure}

\subsubsection*{Mixed measurements}

The same logic applies more generally, but the simple picture of occupied $Z$ bonds forming a loop is no longer sufficient. Once $X$ and $Y$ measurements are included, products of measured operators may combine $X$ and $Z$ components, and stabilizers can convert a seemingly nonpercolating measurement pattern into a logical measurement.

As a first approximation, one may assume a dominant collapse channel. For example, if $p_Z$ is the largest component, the logical collapse is expected to be dominated by percolating logical $Z$ operators, so the detailed values of $p_X$, $p_Y$, and $p_I$ should matter only weakly. This gives the approximate phase boundary $p_Z\simeq0.5$ in the three-dimensional parameter space, in good agreement with the numerical results of Fig.~\ref{fig:numerical_data}.

Pauli-$Y$ measurements require some care. Since $Y_e$ is proportional to $X_eZ_e$, a product of $Y$ measurements on a support $R$ gives, up to an overall phase, $Y_R\sim X_RZ_R$. For this operator to be a toric-code normalizer, the same edge set $R$ must satisfy both the $Z$-logical and $X$-logical normalizer conditions: it must have even degree at every vertex and even overlap with every plaquette boundary. Thus a geometrically percolating cluster of $Y$-measured edges is not automatically a logical operator. The pure-$Y$ line is therefore a {\sl constrained} homological-percolation problem rather than ordinary bond percolation. Consistent with this rigidity of the allowed {$Y$-only} logical supports, our numerics show that the pure-$Y$ transition occurs only at $p_Y=1$ \cite{Lee2026}.

\begin{figure}
    \centering
    \includegraphics[width=1.0\columnwidth]{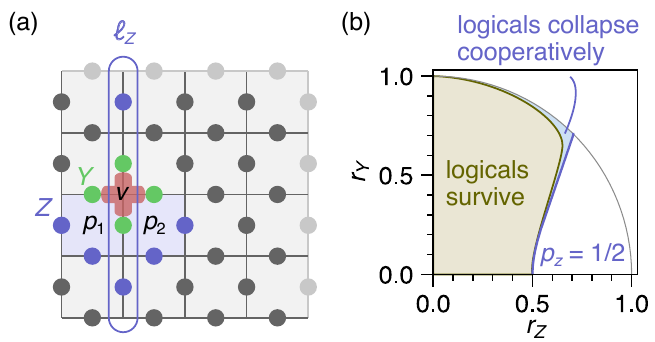}
    \caption{
    {\bf $Y$-assisted logical $Z$ measurement in the toric code.}
    (a) Blue and green markers denote qubits measured in the $Z$ and $Y$ bases, respectively. Up to an irrelevant phase, the product of the corresponding measured Pauli operators is $A_v B_{p_1}B_{p_2}L_Z$, thereby implementing a logical $Z$ measurement, even though the $Z$-measured edges alone support no noncontractible $Z$ loop.
    (b) Schematic $p_X=0$ section of the logical-survival phase diagram. The blue sliver denotes the additional logical-$Z$-measurement regime produced cooperatively by $Y$ and $Z$ measurements below the $Z$-only percolation threshold $p_Z=1/2$, marked by the blue line. Its width is exaggerated for visibility.
    }
    \label{fig:y_assisted_logical_z_measurement}
\end{figure}

At the same time, single-qubit $Y$ measurements can assist the measurement of logical $Z$ or logical $X$ operators even when both $p_Z$ and $p_X$ are below $0.5$. This mechanism explains the (small) deviation of the numerical phase boundary from the simple plane $p_Z=0.5$. An illustrative configuration is shown in Fig.~\ref{fig:y_assisted_logical_z_measurement}.
In this example, the $Z$-measured edges alone support no noncontractible $Z$ loop, while the $Y$ measurements occupy the four edges $\delta v$ incident on a vertex $v$. Up to an overall phase, using $Y_i=iX_iZ_i$, the product of the corresponding measured Pauli operators is
\begin{equation}
  O=X_{\delta v}Z_{M_Z\oplus\delta v} \,,
\end{equation}
where $M_Z$ denotes the $Z$-measured support and $\oplus$ denotes addition of supports modulo two. The product of the corresponding measurement outcomes gives the eigenvalue of $O$.

This picture also connects our phase diagram to the boundary-entanglement problem of Ref.~\cite{Negari2024}, whose fully measured bulk corresponds to the outer surface $r=1$. Within each of the $r_X=0$ and $r_Z=0$ planes, the corresponding outer arc remains in a single short-loop phase away from the pure-$Y$ endpoint~\cite{Nahum2013}. Consistent with this result, the schematic in Fig.~\ref{fig:y_assisted_logical_z_measurement}(b) shows the $Y$-assisted logical-collapse region extending toward the pure-$Y$ point, where its boundary meets the outer surface without introducing an additional transition along the arc. Although the exponentially growing loop correlation length prevents our finite-size numerics from resolving the logical-collapse region all the way to the pure-$Y$ endpoint~\cite{Nahum2013}, we observe a clear deviation from the semi-analytic boundary near the balanced $r_Y=r_Z$ direction.

Along the finite-radius portions of the phase boundary away from the branch junctions, finite-size scaling of the coherent information gives $\nu$ consistent with the two-dimensional percolation value $4/3$ [Fig.~\ref{fig:numerical_data}(d)]. This motivates the common geometric-percolation color used for the inner boundary in Fig.~\ref{fig:phase_diagram}(b).

\subsubsection*{Anyon condensation under measurements}

The toric-code phase diagram Fig.~\ref{fig:phase_diagram}(b) also admits a complementary interpretation in terms of anyon condensation~\cite{Kitaev2003,Eckstein2025}. With the conventions used above, a $Z$ string creates and transports electric charges $e$, whereas an $X$ string creates and transports magnetic fluxes $m$. On each noncontractible cycle, the corresponding closed-string operators are
\begin{equation}
  W_e(\gamma)=\prod_{j\in\gamma}Z_j\equiv Z_L,
  \quad
  W_m(\gamma^*)=\prod_{j\perp\gamma^*}X_j\equiv X_L .
\end{equation}
Thus, when a projective-measurement trajectory fixes $Z_L=\pm1$ or $X_L=\pm1$, the record has resolved an $e$- or $m$-winding sector, respectively.

Once we measure the logical operators $Z_L$ or $X_L$, we obtain the classical bit information of the logical qubit. As a result, the qubit is collapsed into a classical bit. 
From the topological phase perspective, the toric code phase becomes a trivial phase via condensation of $e$ or $m$ anyons, respectively.

This dictionary gives an operational condensation picture for the finite-threshold ``bites'' of the spherical phase diagram. A measured $Z_j$ supplies an elementary $e$-string segment, and products of such measured operators can assemble a noncontractible $W_e$. Once this happens, the record determines the corresponding logical $Z$ eigenvalue and destroys its conjugate $X$ logical coherence. The pure-$Z$ transition at $r_Z=1/2$ may therefore be viewed as proliferation and condensation of $e$ strings. Via electric--magnetic duality, the pure-$X$ transition at $r_X=1/2$ has the same interpretation in terms of $m$ strings. More generally, the inward lobes around the $Z$ and $X$ directions are regimes in which the $e$- or $m$-condensation channel dominates. 
By contrast, the $Y$-dominated regime requires a more careful formulation. Because $Y_j=iX_jZ_j$, products of $Y$ measurements carry electric and magnetic string content simultaneously, which essentially creates a string of the composite $f=e\times m$. However, $f$ is a fermion and therefore cannot be condensed, which can be accounted for by the competition between $e$ and $m$ anyons with nontrivial mutual statistics. Consequently, the $Y$-only logical support must satisfy the direct- and dual-lattice normalizer constraints at the same time, which pushes the pure-$Y$ collapse to the fully measured point $r_Y=1$. The competition between the $e$- and $m$-string sectors thereby provides a useful anyonic explanation for the enhanced robustness around the $Y$ direction.

\section{Color code}
\label{sec:color_code}

We next consider the color code~\cite{Bombin2006} in its standard honeycomb-lattice $(6,6,6)$ incarnation (with periodic boundary conditions). Qubits reside on vertices, and each plaquette $f$ supports both an $X$-type and a $Z$-type stabilizer,
\begin{equation}
  A_f=\prod_{i\in f}X_i,
  \qquad
  B_f=\prod_{i\in f}Z_i .
\end{equation}
The code encodes four logical qubits on the torus. Its geometry and logical-operator conventions are illustrated in Fig.~\ref{fig:schematics}(b,d). 
Because $A_f$ and $B_f$ have identical supports, their product satisfies
\begin{equation}
  A_fB_f\propto\prod_{i\in f}Y_i \,,
\end{equation}
where the overall Pauli phase is irrelevant to the logical-support problem. The stabilizer-support structure is therefore symmetric under permutations of $X$, $Y$, and $Z$. Consequently, the three pure-axis logical-support problems are equivalent, and the phase diagram is invariant under permutations of $(r_X,r_Y,r_Z)$.

\subsubsection*{$Z$-only limit}
As in the toric code, in the pure-$Z$ limit, logical information is lost when the measured qubits support a nontrivial logical $Z$ operator. For the color code, define the qubit--plaquette incidence matrix
\begin{equation}
  D\in\mathbb F_2^{N \times F},
  \quad
  D_{if}=1
  \quad\Longleftrightarrow\quad
  \text{qubit }i\text{ lies on plaquette }f .
\end{equation}
A $Z$-type support $\ell$ commutes with every $X$-plaquette stabilizer when $D^T\ell=0$, while the supports of products of $Z$-plaquette stabilizers form $\operatorname{im}D$. The logical-$Z$ space is therefore
\begin{equation}
  \mathcal L_Z\cong
  \frac{\ker D^T}{\operatorname{im}D},
\end{equation}
where the quotient identifies Pauli operators that differ by stabilizers.

As in the toric code, if $a\in\mathbb F_2^F$ specifies a product of $Z$-plaquette stabilizers, multiplication by that product changes a logical representative according to
\begin{equation}
  \ell\longmapsto\ell+Da .
  \label{eq:color_code_stabilizer_deformation}
\end{equation}
Figure~\ref{fig:schematics}(d) illustrates this operation for the color code. In contrast to Fig.~\ref{fig:schematics}(c), the local moves can split, reconnect, and cancel parts of the logical support rather than merely rerouting a single unbranched path, giving rise to a string-net structure versus a closed-loop structure in the toric code.

The class represented by $\ell$ is measured when one of these representatives is supported entirely in $M$, or equivalently when
\begin{equation}
  D_Ua=\ell_U .
  \label{eq:color_code_du_condition}
\end{equation}
Thus Eqs.~\eqref{eq:toric_code_du_condition} and \eqref{eq:color_code_du_condition} have the same algebraic form. Their physical content differs because the two matrices encode different local relations: a toric-code column is the geometric boundary of a plaquette, whereas a color-code column is the six-qubit support of a plaquette stabilizer. Consequently, the toric-code equivalence reduces to ordinary path deformation, while the color-code equivalence generates branching logical supports that need not contain a single connected path. We refer to the latter as non-geometric homological percolation. The general chain-complex formulation is given in Appendix~\ref{app:homological_percolation}.

Finite-size scaling of the stabilizer-tableau data shown in Fig.~\ref{fig:numerical_data} gives $(r_Z)_{\mathrm c}=0.5000(1)$ and $\nu_{\mathrm{stab}}=1.206(9)$. The threshold is consistent with $1/2$, while the fitted correlation-length exponent differs from the ordinary two-dimensional percolation value $\nu=4/3$.  In fact, the transition point can be fixed analytically. For a CSS code with a single sharp transition in each Pauli sector, the complementary-support relation implies $(r_Z)_{\mathrm c}=1-(r_X)_{\mathrm c}$~\cite{Bravyi2009,Haah2012}. Since the color code is invariant under exchanging $X$ and $Z$, $(r_X)_{\mathrm c}=(r_Z)_{\mathrm c}$, and hence both thresholds must equal $1/2$.

\subsubsection*{Homological connectivity}

The logical-support criterion above is global: it asks whether the measured set supports a nontrivial logical class. To characterize how this global event emerges from local relations, we introduce homological-connectivity classes, in analogy with the clusters used in ordinary geometric percolation. The centers of the plaquettes of each color form one of three interpenetrating triangular sublattices, and the resulting connectivity classes provide the cluster observables used to characterize the transition. Their face-only representation also enables efficient simulations without evolving a full stabilizer tableau.

The local mechanism underlying these classes is stabilizer deformation applied to an open $Z$-type support. Let $q\in\mathbb F_2^N$ denote the support of a $Z$-type Pauli operator whose syndrome consists of two plaquette defects. Multiplication by the $Z$-plaquette stabilizers specified by $a$ changes its support to $q+Da$, as in Eq.~\eqref{eq:color_code_stabilizer_deformation}. Because $D^TD=0$, one has $D^T(q+Da)=D^Tq$ ---the physical support changes while its endpoint defects remain fixed. This is the mechanism by which homological connectivity can extend beyond ordinary path connectivity.

For a qubit $i$, define its syndrome vector with respect to the $X$-plaquette stabilizers by
\begin{equation}
  h_i=(D_{if})_{f=1}^{F}\in\mathbb F_2^F .
\end{equation}
Each $h_i$ has support on the three plaquettes incident on qubit $i$, one of each color. Thus each qubit defines a hyperedge joining three syndrome vertices in the three-uniform hypergraph representation of color-code decoding~\cite{Wang2010,Delfosse2014}. Let $D_M$ denote the restriction of $D$ to the measured qubits. Products of the measured $Z$ operators add their syndrome vectors over $\mathbb F_2$ and generate the reachable-syndrome subspace
\begin{equation}
  R_M=
  \operatorname{span}_{\mathbb F_2}
  \{h_i:i\in M\}
  =\operatorname{im}D_M^T \,.
\end{equation}

Passing to the quotient $\mathbb F_2^F/R_M$ identifies syndrome vectors that differ by an element of $R_M$. Let $e_f$ denote the unit vector associated with plaquette $f$. Two plaquettes $u$ and $v$ of the same color belong to the same homological-connectivity class when
\begin{equation}
  e_u+e_v\in R_M .
  \label{eq:color_code_face_connectivity}
\end{equation}

\begin{figure}
    \centering
    \includegraphics[width=1.0\columnwidth]{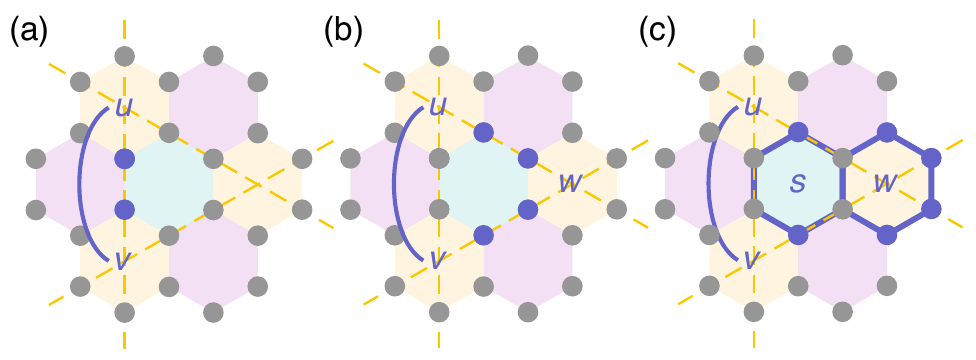}
    \caption{
    {\bf Geometric versus homological connectivity in the color code.}
    $Z$-measured qubits are shown in blue.
    (a) The two measured qubits directly connect the same-color plaquettes $u$ and $v$, giving the direct link $u\sim v$.
    (b) The direct links $u\sim w$ and $w\sim v$ connect $u$ and $v$ through an ordinary geometric path on the same-color triangular sublattice.
    (c) No direct link is present among $u$, $v$, and $w$. Nevertheless, multiplying the measured Pauli product by the two outlined plaquette stabilizers $B_sB_w$ deforms its support into the direct connection shown in (a). Thus $u$ and $v$ are homologically but not geometrically connected.
    }
    \label{fig:cc_hc}
\end{figure}

Figure~\ref{fig:cc_hc} contrasts geometric and homological realizations of this relation. Figure~\ref{fig:cc_hc}(a,b) shows the special case in which direct pair relations form an ordinary path. In Fig.~\ref{fig:cc_hc}(c), stabilizer deformation produces the same relation even though no such path exists. The classes are therefore determined by the full binary span $R_M$, rather than by connected components of the graph formed by direct links. 

To construct these classes, we regard $R_M\subseteq\mathbb F_2^F$ as a classical binary linear code. Gaussian elimination on its generators constructs a parity-check map $\Phi_M$ satisfying $\ker\Phi_M=R_M$. By the coset--syndrome correspondence, $\Phi_M(x)$ labels the coset of any $x\in\mathbb F_2^F$ modulo $R_M$~\cite{Huffman_Pless_2003}. We evaluate this map on the $F$ standard basis vectors and assign each plaquette the label
\begin{equation}
  \lambda_f=\Phi_M(e_f).
\end{equation}
Equivalently, the labels $\lambda_f$ are the columns of a matrix representing $\Phi_M$. By linearity, $\lambda_u=\lambda_v$ precisely when $e_u+e_v\in R_M$. Grouping equal labels within each color therefore gives the homological-connectivity classes~\footnote{When every generator is a pair relation $e_u+e_v$, the resulting classes are ordinary connected components and may be labeled using the Hoshen--Kopelman algorithm~\cite{Hoshen1976}.}.

\begin{figure}
    \centering
    \includegraphics[width=1.0\columnwidth]{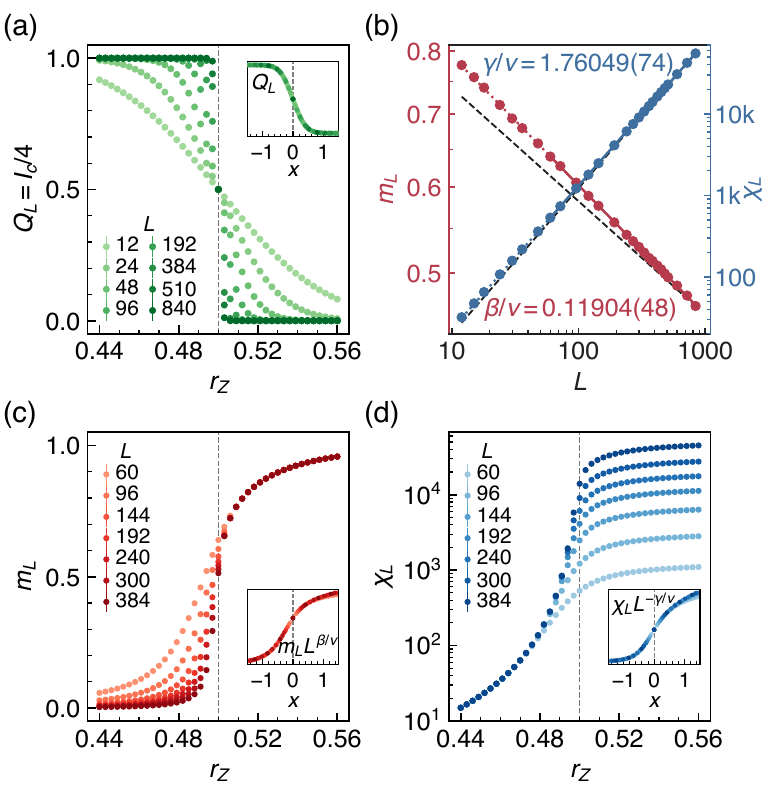}
    \caption{
    {\bf Finite-size scaling of homological percolation.}
    (a) Normalized coherent information $Q_L=I_c/4$, measuring the surviving fraction of the four encoded logical qubits. Critical-slope scaling gives $\nu=1.2073(15)$; the inset shows the corresponding collapse as a function of $x=(r_Z-1/2)L^{1/\nu}$.
    (b) Critical scaling of the largest-class density $m_L$ and the normalized all-class second moment $\chi_L$ of the homological-connectivity classes at $r_Z=1/2$. Fits over $L\geq60$ to $m_L\sim L^{-\beta/\nu}$ and $\chi_L\sim L^{\gamma/\nu}$ give $\beta/\nu=0.11904(48)$ and $\gamma/\nu=1.76049(74)$. Solid, dotted, and black dashed lines denote the fitted range, its continuation to smaller sizes, and the ordinary two-dimensional percolation slopes, respectively.
    (c, d) $m_L$ and $\chi_L$ across the transition; the insets show the corresponding collapses using the exponents from (a) and (b).
    }
    \label{fig:cc_purez_fss}
\end{figure}

Let $V_c$ be the set of plaquettes of color $c$, with $|V_c|=F/3$, and let $s_{j,c}$ be the sizes of the corresponding equivalence classes. We define the largest-class density and normalized all-class second moment as
\begin{equation}
  m_L=
  \frac{1}{3}\sum_c
  \left\langle
    \frac{\max_j s_{j,c}}{|V_c|}
  \right\rangle,
  \qquad
  \chi_L=
  \frac{1}{3}\sum_c
  \left\langle
    \frac{\sum_j s_{j,c}^2}{|V_c|}
  \right\rangle .
  \label{eq:color_code_cluster_observables}
\end{equation}
In particular, $\chi_L$ includes the largest class of each color.

The coherent information is independently obtained from the same face-only linear algebra. The number of independent logical-$Z$ eigenvalues fixed by a realization is
\begin{equation}
  n_{\mathrm{fix}}(M)=
  \bigl(|M|-\operatorname{rank}D_M\bigr)
  -
  \bigl(\operatorname{rank}D-\operatorname{rank}D_U\bigr).
  \label{eq:color_code_logical_count}
\end{equation}
The first term counts products of measured $Z$ operators with vanishing $X$ syndrome, while the second counts those products that are $Z$ stabilizers. Their difference therefore counts the nontrivial logical constraints imposed by the measurement record. Since the periodic color code encodes four qubits, $I_c(M)=4-n_{\mathrm{fix}}(M)$ and $Q_L(p_Z)=\langle I_c(M)\rangle_{p_Z}/4$.

\subsubsection*{Algorithmic improvements}
The homological-connectivity approach not only gives access to the cluster exponents but also permits substantial algorithmic improvements in both memory usage and computational efficiency.
Both the ranks in Eq.~\eqref{eq:color_code_logical_count} and the class labels in Eq.~\eqref{eq:color_code_face_connectivity} depend only on binary supports and their linear relations. They can therefore be computed directly by binary elimination, without maintaining the auxiliary destabilizers used in general tableau simulations. A full stabilizer--destabilizer tableau contains $4N^2$ binary entries to leading order. Omitting destabilizers thus halves this storage. Working in a fixed Pauli sector gives another factor of two, since each operator requires only $N$ support bits rather than $2N$, giving at most $N^2$ entries for a basis. Passing to face-syndrome space reduces both the vector length and the maximum number of independent basis vectors from $N$ to $F=N/2$, reducing this bound to $F^2$ and giving a further factor of four. Finally, storing the basis in echelon form, with the zero entries beyond each pivot omitted, reduces its maximum binary payload to $F^2/2\simeq N^2/8$. Together, these reductions give {\sl approximately $32$ times less storage} than a full tableau for the basis alone. 

The computational savings follow from bit-packed XOR operations and incremental elimination, which reduces each incoming syndrome vector against the existing basis without updating all previously stored rows. For the coherent-information calculation, we additionally add qubits to the measured set one at a time in random order and apply binomial convolution~\cite{Newman2001}, so that each ordering supplies the full probability-dependent curve without separate simulations at each probability. Together, these methods substantially lower the computational and memory cost, permitting $20{,}000$ realizations at each of 26 sizes through $L=840$, corresponding to $N=2L^2=1{,}411{,}200$ physical qubits at the largest size (which is substantially larger than conventional stabilizer tableau calculation which typically top out at around 2,000 to 20,000 qubits depending on circuit depth~\cite{Sierant2022, Klocke2025}).

\subsubsection*{Critical exponents from large-system scaling}

\begin{table}
\caption{{\bf Geometric versus homological percolation.}
Comparison of exact two-dimensional geometric-percolation exponents with the pure-$Z$ color-code estimates at $(r_Z)_c=1/2$. Here $H\equiv2\beta/\nu+\gamma/\nu$; hyperscaling gives $H=d=2$. Parentheses denote sampling standard errors.}
\label{tab:critical_exponents}
\small
\begin{ruledtabular}
\begin{tabular}{lrl}
exponent
& \shortstack{toric-code\\geometric percolation}
& \shortstack{color-code\\homological percolation} \\
\colrule
$\nu$
  & $4/3 \simeq 1.33333 \quad $
  & \qquad $1.2073(15)$ \\
$\beta/\nu$
  & $5/48 \simeq 0.10417 \quad $
  & \qquad $0.11904(48)$ \\
$\gamma/\nu$
  & $43/24 \simeq 1.79167 \quad $
  & \qquad $1.76049(74)$ \\
$\beta$
  & $5/36 \simeq 0.13889 \quad $
  & \qquad $0.14372(62)$ \\
$\gamma$
  & $43/18 \simeq 2.38889 \quad $
  & \qquad $2.1254(27)$ \\[1mm]
\hline\\[-2.5mm]
$H$
  & $2 \quad $
  & \qquad $1.99858(27)$ \\
\end{tabular}
\end{ruledtabular}
\end{table}

Writing $x=(r_Z-1/2)L^{1/\nu}$, the finite-size scaling forms are
\begin{equation}
  Q_L=\mathcal Q(x),
  \qquad
  m_L=L^{-\beta/\nu}\mathcal M(x),
  \qquad
  \chi_L=L^{\gamma/\nu}\mathcal X(x).
  \label{eq:color_code_fss}
\end{equation}
Fixing the exact critical point $(r_Z)_{\mathrm c}=1/2$, a fit of the critical slope $-\left.\partial_{r_Z}Q_L\right|_{r_Z=1/2}\propto L^{1/\nu}$ gives 
\[
	\nu=1.2073(15) 
\] 
for the homological percolation transition.
Critical-point fits over $L\geq60$ determine the remaining exponent ratios, with the fits and scaling collapses shown in Fig.~\ref{fig:cc_purez_fss}, and the resulting exponents summarized in Table~\ref{tab:critical_exponents}.
As a consistency check, we find that the calculated exponents obey hyperscaling 
\footnote{The slight deficit of $H$ relative to $2$ suggests residual finite-size corrections, which can be reduced by restricting the fits to larger system sizes. The clear distinction of the critical exponents from those of ordinary two-dimensional percolation remains untouched also for such a restricted analysis.} $H\equiv2\beta/\nu+\gamma/\nu = d = 2$.
Together, these results sharpen the moderate-size tableau estimate and provide evidence for a distinct stabilizer-defined homological-percolation universality class.

It is natural to expect a common universality class for color codes on different lattice geometries.
Preliminary calculations extending the planar color-code analysis on the (4,8,8) `square-octagon' lattice of Ref.~\cite{Colmenarez2025}
show an effective correlation-length exponent that slowly drifts from a value around $\nu \approx 4/3$ (reported in~\cite{Colmenarez2025} for system sizes $9\leq d\leq25$) towards smaller values clearly distinct from $\nu=4/3$ as considerably larger system sizes are included~\footnote{In complementary simulations with periodic boundary conditions, we also observe distinct effective cluster exponents for the square and octagonal face sectors. This sector dependence is reminiscent of the different sublattice magnetic exponents in the three-spin Union Jack model~\cite{Ding2013}. A related distinction between square and octagonal plaquettes appears in the vison correlations of a square-octagon quantum dimer model~\cite{Shah2026}.}.
We leave a systematic investigation of the observed drift in our setting to future work.

\subsubsection*{Mixed measurements}

By the stabilizer-support symmetry established above, the pure-$X$, pure-$Y$, and pure-$Z$ color-code problems are equivalent. All three transitions therefore occur at measurement probability $1/2$ and share the same critical behavior.

Treating the three logical Pauli sectors independently gives the semi-analytic phase boundary shown in Fig.~\ref{fig:phase_diagram}(c). The numerically determined logical-survival boundary broadly follows this construction. Away from the regions where two collapse branches meet, stabilizer-tableau scaling along representative mixed $X/Z$ cuts is also consistent with $\nu\simeq1.2$.

\begin{figure*}
    \centering
    \includegraphics[width=1.0\textwidth]{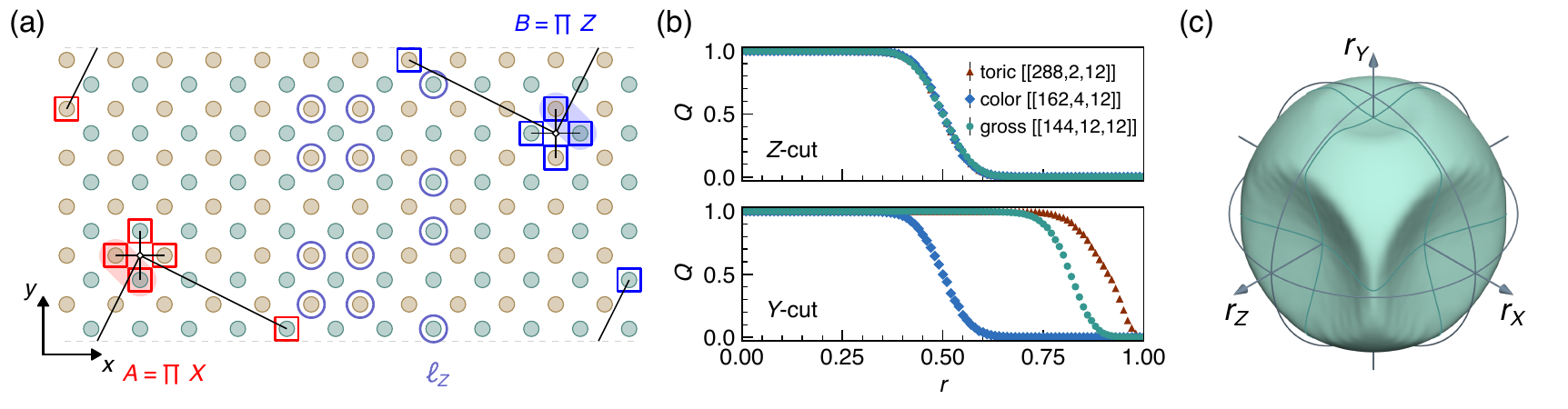}
    \caption{
    {\bf Structure and logical survival of the finite gross code.}
    (a) Periodic $12\times12$ physical-qubit layout of the gross $[[144,12,12]]$ code. Red and blue squares show the supports of representative $X$- and $Z$-type stabilizers $A$ and $B$, respectively, while the red and blue ellipses identify their reference cells. Black lines show the six check--data couplers of each stabilizer in the layout of Ref.~\cite{Yoder2025}. The qubits encircled in purple form one representative $\ell_Z$ of a logical $Z$ operator.
    (b) Normalized coherent information $Q=\langle I_c\rangle/k$ under pure-$Z$ and pure-$Y$ measurements for the distance-matched toric $[[288,2,12]]$, color $[[162,4,12]]$, and gross $[[144,12,12]]$ codes. In the upper and lower plots, $r=r_Z$ and $r=r_Y$, respectively, with all other measurement probabilities set to zero.
    (c) Finite-code survival diagram of the gross code. The mint surface is the boundary where half of the encoded information survives, $Q=1/2$. It is obtained numerically from radial scans along a grid of directions.
    }
    \label{fig:gross_code}
\end{figure*}

\subsubsection*{Anyon condensation under measurements}

The color code has a richer anyon content compared with the toric code~\cite{Kesselring2024}. At the level of anyon theory, the color code admits a parent description with three copies of color-labelled toric codes with the composite electric boson $e_r e_g e_b$ condensed~\cite{Davydova2026}. Consequently, the condensation $e_r e_g e_b=1$ leads to two independent copies of toric codes while retaining three colored string species that can meet at trivalent junctions. 
Specifically, the color code has 16 anyons -- the vacuum $1$, nine bosons $cp$ labeled by a color $c\in\{r,g,b\}$ and a Pauli index $p\in\{x,y,z\}$, and six fermions. For each fixed $p$, $\mathcal L_p=\{1,rp,gp,bp\}$ is a Lagrangian subgroup -- the maximal set of mutually bosonic anyons. In particular, every anyon outside the Lagrangian subgroup braids nontrivially with at least one element in it. Consequently, all the anyons in the Lagrangian subgroup can be condensed simultaneously.  

In the random measurement problem, a homologically nontrivial measured string-net fixes the corresponding logical operator and destroys its conjugate logical coherence. The three ``bites'' in Fig.~\ref{fig:phase_diagram}(c) are the measurement analogues of proliferating $\mathcal L_x$, $\mathcal L_y$, and $\mathcal L_z$, which are permuted by permutations of the Pauli matrices. The purple color, on the other hand, indicates the phase where the condensations of the three Lagrangian subgroups compete with each other, leaving the system in a critical state~\footnote{The competition between Lagrangian subgroups in 2D color code is similar to that in the 2$d$ toric code between $e$ and $m$ condensations. In particular, in the toric code, the $e$-$m$ competition can be equivalently described by the competition between their respective Lagrangian subgroups $\mathcal{L}_e = \{ 1,e \}$ and $\mathcal{L}_m = \{1, m\}$.}.

\section{Gross code}

Having established the distinct character of geometric versus homological percolation in the principal examples of the toric code and color code, 
we next turn to another family of CSS codes --  bivariate bicycle (BB) codes~\cite{MacKay2004,Bravyi2024} and, 
in particular, the gross code~\cite{Bravyi2024} which has attracted increasing attention in proposals for scalable 
quantum error-correcting architectures~\cite{Yoder2025}.
Because of its practical relevance, we will mostly discuss the {\sl finite} $[[144,12,12]]$ gross code (with a dozen dozen, i.e., a gross of physical qubits) 
in the following, and only briefly touch
on the large code distance (thermodynamic) limit.

\subsubsection*{Bivariate bicycle (BB) codes}
Let us set the stage by briefly introducing BB codes and their topological and logical properties. Let $\Lambda=\mathbb Z_\ell\times\mathbb Z_m$ be a periodic square lattice. Each cell $\bm r\in\Lambda$ contains two data qubits, denoted $h$ and $v$, so that the total number of physical qubits is $N=2|\Lambda|=2\ell m$. We use a polynomial convention to specify the stabilizer supports~\cite{Haah2013},

\begin{equation}
  \begin{aligned}
    f(x,y)&=1+x+x^{-\alpha}y^b,\\
    g(x,y)&=1+y+x^ay^{-\beta}.
  \end{aligned}
\end{equation}
The stabilizer code defined by such a pair $(f,g)$ is termed a $\mathbb{BB}(\bar\alpha,\bar\beta,a,b)$ code, where $\bar\alpha\equiv-\alpha$ and $\bar\beta\equiv-\beta$. Here $x$ and $y$ generate translations on $\Lambda$, and the $x$- and $y$-coordinates are understood modulo $\ell$ and $m$, respectively.

For a polynomial $p$, define $\bar p(x,y)=p(x^{-1},y^{-1})$. In this notation, the $X$- and $Z$-type stabilizer support maps are represented as
\begin{equation}
  D_X=
  \begin{bmatrix}
    f\\
    g
  \end{bmatrix},
  \qquad
  D_Z=
  \begin{bmatrix}
    \bar g\\
    \bar f
  \end{bmatrix},
\end{equation}
where the first (second) component represents the positions of the $h$ ($v$) qubits. Expanding the translations gives $N\times|\Lambda|$ binary support matrices, which satisfy $D_X^TD_Z=0$, as required for the $X$- and $Z$-type stabilizers to commute.

The gross code is the $\mathbb{BB}(\bar{1},\bar{1},3,3)$ code on the $(\ell,m)=(12,6)$ lattice. In the convention used here,
\begin{equation}
  f=1+x+x^{-1}y^3,
  \quad
  g=1+y+x^3y^{-1},
  \quad
  x^{12}=y^6=1.
\end{equation}
Its $12\times6$ cells, each containing two data qubits, are displayed as the periodic $12\times12$ physical-qubit layout in Fig.~\ref{fig:gross_code}(a). For the representative $X$ stabilizer $A$, the polynomials $f$ and $g$ specify its support on the yellow $h$ and green $v$ qubits, respectively; the representative $Z$ stabilizer $B$ is similarly specified by $\bar g$ and $\bar f$. The pale red and blue ellipses identify their reference cells. The qubits encircled in purple form one representative $\ell_Z$ of a logical $Z$ operator.

On the torus, the logical-$Z$ and logical-$X$ support spaces are
\begin{equation}
  \mathcal L_Z\cong
  \frac{\ker D_X^T}{\operatorname{im}D_Z},
  \qquad
  \mathcal L_X\cong
  \frac{\ker D_Z^T}{\operatorname{im}D_X},
\end{equation}
respectively. As before, these quotients identify Pauli supports that differ only by stabilizers. The number of logical qubits generically depends on both the BB code parameters $(\bar{\alpha},\bar{\beta},a,b)$ and the system size $(\ell,m)$. 
For instance, for the $\mathbb{BB}(\bar{1},\bar{1},3,3)$ code, the number of encoded logical qubits is~\cite{Chen2025}
\begin{equation}
  k(\ell,m)=
  \begin{cases}
    16,&12\mid q,\\
    12,&6\mid q,\;12\nmid q,\\
    8,&3\mid q,\;6\nmid q,\\
    0,&3\nmid q,
  \end{cases}, \quad q = \text{gcd}(\ell,m).
\end{equation} 
For $(\ell,m)=(12,6)$, this gives $k=12$ and hence the $[[144,12,12]]$ gross code shown in Fig.~\ref{fig:gross_code}(a).

\subsubsection*{Random measurements on BB codes}
We now consider random measurements on the finite $[[144,12,12]]$ gross  code. Figure~\ref{fig:gross_code}(b) compares the normalized coherent information of the gross, toric, and color codes at the same code distance along the pure-$Z$ and pure-$Y$ axes. At this distance, the three pure-$Z$ curves nearly coincide. Along the pure-$Y$ axis, the color-code response remains identical to its pure-$Z$ response, whereas the toric and gross codes retain logical information to substantially larger measurement probabilities. The gross-code curve lies between the color- and toric-code curves, although closer to the latter.

In the finite-code phase diagram of Fig.~\ref{fig:gross_code}(c), logical information survives to higher measurement probabilities near the pure-$Y$ direction. Calculations for several larger tori suggest that the small indentation around this direction is a finite-size feature that disappears in the large code-distance (thermodynamic) limit, which would make the gross-code phase diagram indistinguishable from the toric-code phase diagram.

The gross code is one finite member of a BB family whose anyon content generalizes the branching logical-support structure encountered in the color code~\cite{Kesselring2024,Davydova2026}. Its larger collection of toric-code sectors provides a richer set of anyon labels, and its logical representatives can form branched networks of anyon-labelled strings with junctions constrained by fusion rules. Motivated by this observation, we examine the pure-$Z$ transition along three scaling sequences of the $\mathbb{BB}(\bar{1},\bar{1},3,3)$ family, with $k=8$, $12$, and $16$, which differ in their aspect ratios and translation sectors. The complementary-support relation and $X/Z$ symmetry fix the transition at $(r_Z)_{\mathrm c}=1/2$.

\begin{figure}[b]
    \centering
    \includegraphics[width=1.0\columnwidth]{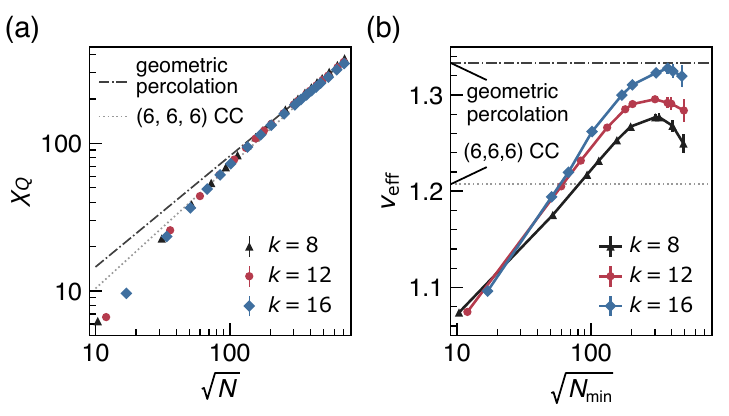}
    \caption{{\bf Finite-size scaling of the BB-code family under pure-$Z$ measurements.} Results for three scaling sequences of the $\mathbb{BB}(\bar{1},\bar{1},3,3)$ family with $k=8$, $12$, and $16$. (a) Critical slope $\chi_Q=-\left.\partial_{r_Z}Q\right|_{r_Z=1/2}$ versus $\sqrt{N}$. (b) Effective exponent $\nu_{\mathrm{eff}}$ obtained by fitting $\chi_Q\propto N^{1/(2\nu_{\mathrm{eff}})}$ over sizes $N\geq N_{\min}$. }
    \label{fig:bb_family_scaling}
\end{figure}

The lower-cutoff dependence of the finite-size fits is shown in Fig.~\ref{fig:bb_family_scaling}. Strong window dependence persists up to approximately $5\times10^5$ physical qubits. As the smallest sizes are omitted, the fit quality improves and the estimates initially move toward the geometric-percolation value $\nu=4/3$. At the largest lower cutoffs, however, the central estimates turn downward in all three sequences. This trend suggests a universality class distinct from ordinary geometric percolation, although the absence of a common plateau leaves the thermodynamic value of $\nu$ unresolved.

The exceptionally long drift may reflect the length scale associated with anyon motion. Translating an isolated defect by one lattice spacing can change its anyon type, so transporting the same charge generally requires intermediate configurations containing several defects. In this BB family, translations by twelve lattice spacings along either axis preserve all anyon types, defining a mobility sublattice with a $12\times12$ unit cell~\cite{Chen2025}. Even our largest lattices therefore contain far fewer unit cells of this mobility sublattice than microscopic cells. Together with the multi-species fusion structure of the logical string networks, this additional length scale may contribute to the slow approach to asymptotic scaling.

\section{General QEC codes}
\label{sec:wider_perspective}

Let us now go beyond the three CSS examples in two spatial dimensions which we have discussed so far and show that random-measurement transitions can share a homological-percolation description even when their critical behavior differs. To this end, we extend our description to more general QEC codes, using the anyon-condensation picture to help interpret their phase diagrams. In particular, we study the Wen plaquette model (as a representative non-CSS code), 3$d$ toric and color codes, subsystem codes, and Floquet codes, with the results summarized in Fig.~\ref{fig:family}.

We then connect this measurement problem to error correction under decoherence and logical inference from weak measurements. These connections give bounds on decoding and learning thresholds, while a protocol combining recorded measurements with hidden errors makes the relation between homological percolation and decoding explicit.

\subsection{Random measurements}
\label{sec:general_qec_codes}

\begin{figure}[t]
    \centering
    \includegraphics[width=1.0\columnwidth]{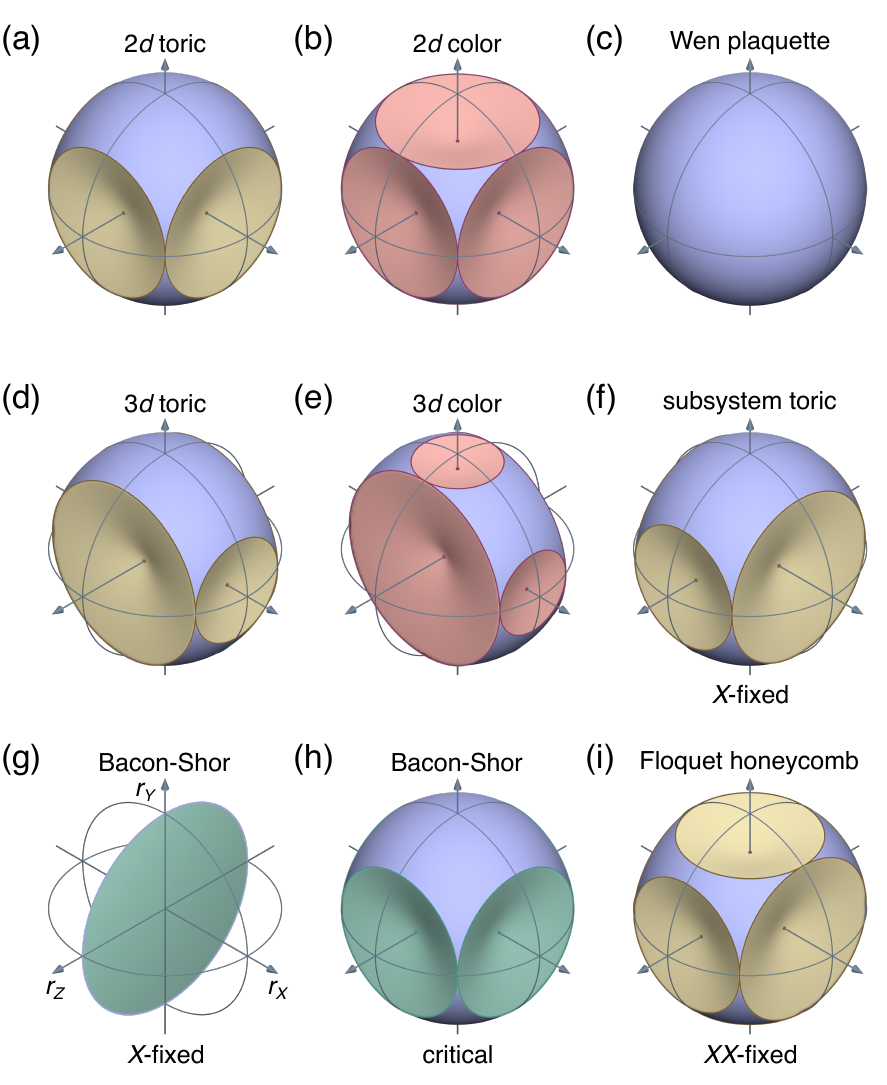}
    \caption{
    {\bf Semi-analytic phase diagrams of randomly monitored QEC codes.}
    Numerically determined thresholds along the pure-$X$, pure-$Y$, and pure-$Z$ axes are used to construct the approximate phase boundaries. Surface colors indicate the critical behavior: yellow for geometric percolation, red for non-geometric homological percolation, and mint for critical behavior outside these two categories. (a) 2$d$ toric code; (b) 2$d$ color code; (c) Wen plaquette model; (d) 3$d$ toric code; (e) 3$d$ color code; (f) $X$-fixed subsystem toric code; (g) $X$-fixed Bacon--Shor code; (h) critical steady-state Bacon--Shor code; and (i) the post-$XX$ instantaneous stabilizer group of the Floquet honeycomb code.
    }
    \label{fig:family}
\end{figure}

We continue to consider the random single-qubit Pauli measurement protocol, and the central question is unchanged---for each realization, we ask whether products of the measured operators yield a nontrivial logical Pauli operator. This gives the general stabilizer-code formulation of homological percolation, described in Appendix~\ref{app:homological_percolation}. All such instances can be simulated efficiently within the Clifford stabilizer formalism. Figure~\ref{fig:family} surveys a range of two- and three-dimensional examples. The phase boundaries are constructed semi-analytically from the pure-axis thresholds and generally agree well with the numerical results, with the Floquet honeycomb code as the main exception. Across these code families, the phase diagrams vary markedly, and finite-size scaling gives evidence that their transitions can belong to distinct universality classes. Representative thresholds and scaling estimates are collected in Table~\ref{tab:family_scaling}. Below, we discuss the structure of these phase diagrams and their dependence on the code and its preparation.

\begin{table*}
\caption{{\bf Critical scaling across randomly monitored QEC codes.}
Pure-$Z$ thresholds and estimates of $\nu$ for selected models in Fig.~\ref{fig:family}. Except for the 2$d$ color-code result, which is obtained from the dedicated face-only calculation, the numerical data are obtained from Clifford stabilizer simulations. Arrows show the drift when smaller sizes are excluded. Parentheses denote sampling standard errors; systematic uncertainties associated with the fitting ansatz and residual finite-size effects are not included.
}
\label{tab:family_scaling}
\small
\begin{ruledtabular}
\begin{tabular}{llcl}
model
& \multicolumn{1}{c}{$(r_Z)_{\mathrm c}$}
& sizes $L_{\min}$--$L_{\max}$
& $\nu$ \\[0.5mm]
\colrule\\[-2.5mm]
2$d$ toric code
& \multicolumn{1}{c}{$1/2$}
& $36$--$120$
& $1.334(43)$ \\
2$d$ color code
& \multicolumn{1}{c}{$1/2$}
& $60$--$840$
& $1.2073(15)$ \\
3$d$ toric code
& $0.24885(14)$
& $14$--$24$
& $0.8659(80)$ \\
3$d$ color code
& $0.161160(13)$
& $8$--$16\rightarrow12$--$16$
& $0.66925(61)\rightarrow0.6879(22)$ \\
Subsystem toric code ($X$-fixed)
& $0.65288(24)$
& $28$--$64$
& $1.325(13)$ \\
Floquet honeycomb code (post-$XX$ ISG)
& $0.58972(43)$
& $42$--$72$
& $1.381(24)$ \\
Bacon--Shor code (critical-prepared)
& $0.5057(48)$
& $16$--$56$
& $5.76(25)$ \\
\end{tabular}
\end{ruledtabular}
\end{table*}

\subsubsection*{Wen plaquette model}

The Wen plaquette model gives a non-CSS realization of the same topological order as the toric code~\cite{Wen2003}. Qubits are put on the vertices of an $L\times L$ square lattice with even $L$ and periodic boundary conditions. The stabilizers are
\begin{equation}
  W_{x,y}=X_{x,y}Z_{x+1,y}X_{x+1,y+1}Z_{x,y+1}.
\end{equation}
These commuting plaquette operators define a code with two logical qubits. Applying Hadamard gates on one checkerboard sublattice maps the model to the CSS toric code. This transformation also exchanges physical $X$ and $Z$ measurements on that sublattice, so a spatially uniform measurement protocol in one representation generally becomes sublattice dependent in the other~\cite{Negari2024}.

Plaquette defects on the two checkerboard sublattices represent $e$ and $m$ excitations. A local $X$ or $Z$ operator creates a pair of either type, depending on the sublattice. The model is particularly robust to strongly biased noise~\cite{BonillaAtaides2021}. Under pure $Z$ noise, both types of defects move only along parallel diagonals within their respective sublattices, reducing the decoding problem to a collection of 1$d$ repetition codes.

Likewise, along the pure-$Z$ measurement axis of our protocol, logical collapse requires a fully measured diagonal logical string, giving the threshold $(r_Z)_{\mathrm c}=1$. The same argument applies to pure-$X$ measurements along the anti-diagonal. Since Hadamard gates preserve the $Y$ measurement basis, the pure-$Y$ protocol is equivalent to that of the toric code and likewise has threshold $(r_Y)_{\mathrm c}=1$. All three pure axes therefore have unit thresholds, yielding a semi-analytic phase diagram with no inward bites, as shown in Fig.~\ref{fig:family}(c). Numerical scans over mixed measurement directions support this phase diagram.

\subsubsection*{Three-dimensional topological codes}

Next we consider two three-dimensional CSS topological codes---the 3$d$ toric code~\cite{Hamma2005,Castelnovo2008} and the 3$d$ color code~\cite{Bombin2007,Kubica2018}. One important change from the two-dimensional case is the nature of the logical operators, which now include both one-dimensional strings and two-dimensional membranes.

For the 3$d$ toric code, the $Z$-type logical operators are noncontractible strings. An open $Z$ string creates a pair of pointlike $e$ excitations at its endpoints, while closing the string removes these excitations. In the pure-$Z$ measurement limit, logical collapse therefore occurs when the measured edges contain a nontrivial winding cycle. This is ordinary bond percolation on the cubic lattice, reflecting the identification of the toric-code chain complex with the cellular chain complex of that lattice. As shown in Table~\ref{tab:family_scaling}, the estimated threshold and critical exponent agree with the known three-dimensional bond-percolation values~\cite{Wang2013}.

At first glance, the pure-$X$ limit presents a dimensional mismatch, since we measure qubits on edges while the logical operators are membranes. On the dual cubic lattice, however, each edge corresponds to a plaquette, so the problem becomes geometric homological percolation of closed membranes~\cite{Duncan2023}. The complementary relation of the CSS code gives $(r_X)_{\mathrm c}=1-(r_Z)_{\mathrm c}$ and the same critical exponent for both transitions. Unlike the square-lattice toric code in two dimensions, this model has no $X/Z$ self-duality, and its thresholds satisfy $(r_X)_{\mathrm c}>(r_Z)_{\mathrm c}$. From the excitation perspective, the pure-$X$ transition involves membranes bounded by flux loops, in contrast to the pointlike charges associated with $Z$ strings. As in the two-dimensional toric code, the additional constraints on $Y$-only logical supports suppress logical collapse. Our numerical results are consistent with a transition only in the fully measured limit $(r_Y)_{\mathrm c}=1$.

Like the 2$d$ color code, the 3$d$ color code admits branching logical supports. Its $Z$ strings carry four possible colors, and a string of one color can split into three strings of the other colors, forming a string-net. Similarly, $X$ membranes are labeled by pairs of colors and can join into membrane-nets~\cite{Bombin2007}. Their measurement transitions again define homological-percolation problems, with the pure-$X$ and pure-$Z$ boundaries related by CSS complementarity. Within the system sizes studied, however, we cannot yet resolve the thermodynamic value of the critical exponent $\nu$. As in the BB-code family, the estimates show sizable drift when we restrict the fits to larger systems---we leave the determination of the asymptotic universality class for future work.

Qualitatively, the phase diagram of the 3$d$ color code resembles that of its two-dimensional counterpart, with bites along all three pure axes. The same stabilizer group can be generated using either $X$-type or $Y$-type operators together with the $Z$-type stabilizers, since the corresponding $X$- and $Y$-type generators differ only by products of $Z$-type stabilizers~\cite{Bombin2007,Kubica2015}. This gives an exact $X/Y$ symmetry of the measurement protocol, so the pure-$X$ and pure-$Y$ thresholds coincide.

\subsubsection*{Subsystem and Floquet codes}

Finally, we consider subsystem codes~\cite{Poulin2005,Bombin2010} and Floquet codes~\cite{Hastings2021,Haah2022}. Unlike the stabilizer subspace codes considered above, these require an additional choice to specify the initial logical subspace. For subsystem codes, we fix the state of the gauge qubits, while for Floquet codes, we choose a round of the measurement cycle. Once this choice is made, the resulting stabilizer group defines a subspace code to which our homological-percolation formulation applies, as described in Appendix~\ref{app:homological_percolation}. Random projective measurements of subsystem codes were also studied in Ref.~\cite{Lee2026}, with the gauge qubits initially maximally mixed. Here, we instead fix their state.

We investigate three subsystem-code preparations. First, we consider the subsystem toric code~\cite{Bravyi2013}, with its gauge qubits fixed in the $X$ basis. This gauge fixing gives a toric code on the honeycomb lattice~\cite{Higgott2021}, so the same anyon-condensation picture applies. The pure-$Z$ and pure-$X$ transitions correspond to bond percolation on the honeycomb lattice and its triangular dual, respectively. The resulting phase diagram, shown in Fig.~\ref{fig:family}(f), resembles that of the square-lattice toric code but lacks its $X$--$Z$ symmetry. Our numerical thresholds and pure-$Z$ scaling estimates are consistent with these geometric-percolation transitions.

Another example is the Bacon--Shor code~\cite{Shor1995,Bacon2006}. This code is not intrinsically topological, and its measurement response depends strongly on the gauge preparation. We consider two preparations. In the first, we fix all local $XX$ gauge checks, giving Shor's code in the $X$ basis. The logical subspace then collapses for any $r_X>0$ in the thermodynamic limit, as shown in Fig.~\ref{fig:family}(g). In the second, we prepare the state by randomly measuring local $XX$ and $ZZ$ check operators at equal rates, retaining the preparation outcomes~\cite{Lavasani2021,Sharma2023,Placke2025}. The corresponding steady-state ensemble is critical, with bipartite entanglement entropy scaling as $L\ln L$~\cite{Sharma2023}. For this critical preparation, the phase diagram in Fig.~\ref{fig:family}(h) resembles that of the toric code. However, finite-size scaling along the pure-$Z$ axis gives an unusually large exponent $\nu=5.76(25)$, suggesting behavior beyond ordinary geometric percolation. This opens an opportunity to explore homological percolation on a critical background, in a similar spirit to percolation on critical Ising configurations~\cite{Chen2025_percolation} and recursive percolation, where large correlation-length exponents have also been reported~\cite{Liu2015}.

For the Floquet honeycomb code, we consider the variant in which the three edge colors of the honeycomb lattice form a Kekul\'e pattern and are identified with the $XX$, $YY$, and $ZZ$ bond operators~\cite{Gidney2021}. The toric-code phase is maintained by sequentially measuring all bonds of each type. We choose the initial logical subspace defined by the instantaneous stabilizer group (ISG) immediately after the $XX$ round, giving a measurement phase diagram symmetric under exchanging $Y$ and $Z$, as reflected in the semi-analytic boundary in Fig.~\ref{fig:family}(i).
The numerical results, however, show a substantial deviation from this prediction for mixed $Y$ and $Z$ measurements, which we discuss using the toric-code description in Appendix~\ref{app:Floquet}. Finite-size scaling of these transitions is broadly consistent with two-dimensional geometric percolation, although the estimated exponents depend somewhat on the fitting method and system-size range.

\subsection{Monitoring, decoding, and inference transition}

Let us now step back and discuss how the random-monitoring protocol and the phase diagrams calculated for it provide quantitative and qualitative guidance for two closely connected problems -- the decoding transition in the presence of incoherent noise, and the inference/learning transition for weak single-qubit measurements.

\subsubsection*{Decoding threshold bounds}

In the random-monitoring protocol, we retain the complete record of every local projective measurement---whether a measurement occurred and, if so, its outcome.  A recorded single-qubit $Z$ measurement applied with probability $r$ is therefore described by the quantum channel~\cite{Nielsen2010}
\begin{equation}
  \begin{aligned}
    \mathcal{F}_{r}^{Z}(\rho) ={}&(1-r) \bigl(\ket{\mathrm{u}}\!\bra{\mathrm{u}}\bigr)_{C} \otimes\rho \\
    &+r\sum_{a=0,1} \bigl(\ket{\mathrm{m},a}\!\bra{\mathrm{m},a}\bigr)_{C} \otimes P_a^Z\rho P_a^Z ,
  \end{aligned}
\end{equation}
where $C$ is a three-state classical register and $P_a^Z=[I+(-1)^aZ]/2$. Discarding $C$ forgets both the realized measurement pattern and its outcomes, and yields a {\sl dephasing channel} with phase-flip probability $q=r/2$
\begin{equation}
  \begin{aligned}
    \operatorname{Tr}_{C}\!\left[\mathcal{F}_{r}^{Z}(\rho)\right] &=(1-r)\rho+r\sum_{a=0}^{1}P_a^Z\rho P_a^Z \\
    &=(1-q)\rho+qZ\rho Z\equiv\mathcal{D}_{q}^{Z}(\rho).
  \end{aligned}
\end{equation}
Since a recovery with access to $C$ can always discard it before decoding, for a given code,
\begin{equation}
  r_\mathrm{c}^Z\geq(r_\mathrm{c}^Z)^{\mathrm{forgotten}}=2(q_\mathrm{c}^Z)^{\mathrm{ML}},
\end{equation}
where $\mathrm{ML}$ denotes maximum-likelihood decoding.

One can also obtain a lower bound on the same dephasing threshold following a Peierls counting argument~\cite{Dennis2002, Dumer2015}. Let $A_L(\ell)$ count irreducible syndrome-free supports of weight $\ell$ that represent a nontrivial $Z$-logical class, i.e., the essential supports of logical $Z$ operators after detached stabilizer-trivial pieces have been removed. In the toric code, for example, this excludes the decorated representative $\ell_Z+B_p$ when $B_p$ is an isolated plaquette loop disjoint from the winding logical string $\ell_Z$. If $A_L(\ell)\lesssim\operatorname{poly}(L)\mu_{\mathrm{log}}^\ell$ (with $d_Z(L)\gg\log L$, as for the codes considered here), following the Peierls counting argument, the threshold for (global) minimum-weight decoding is {\sl lower-bounded} as
\begin{equation}
  \begin{aligned}
  (q_\mathrm{c}^Z)^{\mathrm{ML}}
  &\geq (q_\mathrm{c}^Z)^{\mathrm{MW}}
  \geq q_{\mathrm{P}}(\mu_{\mathrm{log}}),\\
  q_{\mathrm{P}}(\mu)
  &=\frac{1-\sqrt{1-\mu^{-2}}}{2}.
  \end{aligned}
\end{equation}
Intuitively, $\mu_{\mathrm{log}}$ counts the exponential growth of the number of distinct irreducible logical $Z$ operators with their weight. Minimum-weight decoding can fail only if there is an accidental logical support of weight $\ell$ on which at least half of the qubits are affected by dephasing errors. Therefore, a larger $\mu_{\mathrm{log}}$ increases the number of possible failure modes for the decoder. The value of $\mu_{\mathrm{log}}$ is controlled by the underlying chain complex, which determines the syndrome-free supports, the trivial boundaries, and the allowed deformations shown in Fig.~\ref{fig:schematics}. For the toric code, this is the growth of nontrivial simple loops on the square lattice, whereas for the color code the corresponding objects are colored branching string nets.

\begin{figure}
    \centering
    \includegraphics[width=1.0\columnwidth]{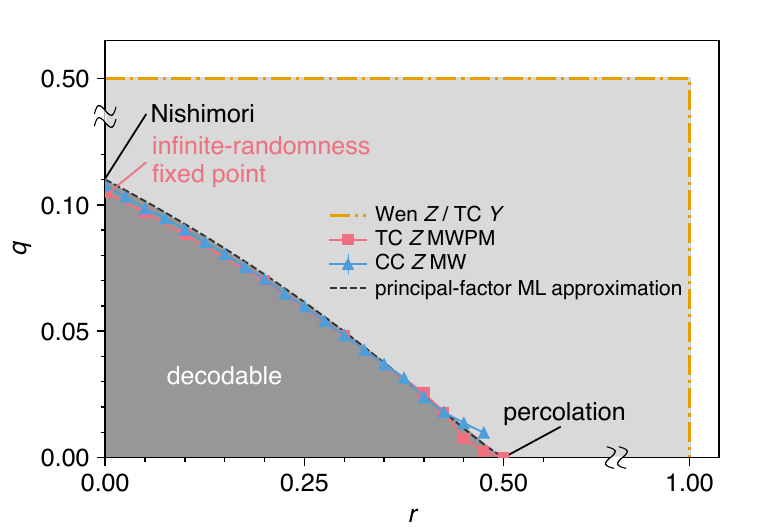}
    \caption{{\bf Measurement--decoherence phase diagram.} 
    The Wen-plaquette $Z/Z$ and toric-code $Y/Y$ protocols attain the maximal noise threshold $q_c=1/2$ for all $r<1$, with the decodable region shaded light gray. 
    The toric-code $Z/Z$ data are taken from the minimum-weight perfect-matching (MWPM) calculation~\cite{Stace2009}, whose maximal threshold theory (in the absence of any random monitoring, $r=0$) has recently been identified as an infinite-randomness fixed point \cite{Pandey2026}.
    For the $(6,6,6)$ color code, minimum-weight (MW) decoding estimates are obtained by extrapolating the $P_{\mathrm{full}}=1/2$ contours using $L=9,12,15$ for $r<0.4$ and $L=18,21,24$ for $r\geq0.4$. 
    The MW decoding estimates already lie close to the principal-factor approximation to the maximum-likelihood (ML) threshold, with dark gray shading indicating the decodable region. 
    }
    \label{fig:measurement_vs_decoherence}
\end{figure}

\subsubsection*{Measurement--decoherence phase diagram}

To make the connection between homological percolation and error correction under decoherence explicit, we introduce a two-parameter extension of our measurement protocol. We consider aligned $Z/Z$ and $Y/Y$ protocols, in which each physical qubit is projectively measured in the chosen Pauli basis with probability $r$, followed by an independent, unobserved error of the same Pauli type with probability $q$. The measurement locations and outcomes remain available to the decoder, and we assume perfect final syndrome extraction. This construction continuously interpolates between the two protocols---at $q=0$, it reduces to the homological-percolation problem, while at $r=0$, it becomes the usual decoding problem for a code subjected to dephasing in the chosen basis.

For either choice $\sigma\in\{Z,Y\}$, the measurement projector satisfies $\sigma P_a^\sigma=(-1)^aP_a^\sigma$, and we obtain
\begin{equation}
  \mathcal{D}_q^\sigma(P_a^\sigma\rho P_a^\sigma)=P_a^\sigma\mathcal{D}_q^\sigma(\rho)P_a^\sigma=P_a^\sigma\rho P_a^\sigma.
\end{equation}
Thus, dephasing before or after the measurement gives the same final density matrix for each recorded outcome, and errors of the chosen Pauli type on measured qubits have no effect.

As discussed above, for CSS codes with electric--magnetic symmetry, recorded $Z$ measurements and detectable qubit loss have the same threshold. This correspondence extends to additional hidden dephasing, with both settings described by diluted generalized Ising models with random couplings, including the random-bond Ising model for the toric code and the random three-body Ising model for the color code~\cite{Stace2009, Stace2010, Ohzeki2012,Colmenarez2025, English2026}.

Figure~\ref{fig:measurement_vs_decoherence} summarizes the phase diagram for the selected models. For the Wen-plaquette code, decoding under pure $Z$ noise reduces to diagonal repetition codes with threshold $q_c=1/2$, underlying its robustness to strongly biased noise~\cite{BonillaAtaides2021}. Recorded $Z$ measurements combine neighboring syndrome checks, leaving repetition chains whose lengths still grow with system size for any fixed $r<1$. The toric code likewise has the maximal threshold under pure $Y$ noise~\cite{Tuckett2019}. A common argument establishes the mixed threshold in both cases. Discarding the measurement record gives an effective Pauli-error probability $q_{\mathrm{eff}}=r/2+(1-r)q$, which remains below $1/2$ whenever $r<1$ and $q<1/2$. Since retaining the record cannot reduce recoverability, both aligned protocols attain $q_c=1/2$ for all $r<1$.

For the toric-code $Z/Z$ protocol, the optimal decoding boundary connects the Nishimori point at $r=0$ to the geometric-percolation endpoint at $(r,q)=(1/2,0)$. The corresponding problem of detectable qubit loss and computational errors was studied using minimum-weight perfect matching~\cite{Stace2009}. These MWPM results, reproduced in Fig.~\ref{fig:measurement_vs_decoherence}, show decreasing noise tolerance as the measurement probability increases, with residual finite-size effects near the percolation endpoint. Note that, in the absence of random monitoring ($r=0$), the MWPM threshold sits slightly below the Nishimori transition and is described by an infinite-randomness fixed point \cite{Pandey2026}.

For the color-code $Z/Z$ protocol, the analogous boundary connects a threshold sitting slightly below the Nishimori point of the random three-body Ising model at $r=0$ to the homological-percolation endpoint at $(r,q)=(1/2,0)$, with the intermediate regime described by diluted random three-body interactions. Unlike toric-code MWPM, exact minimum-weight decoding of color codes is NP-hard in general~\cite{Walters2026}. Following the MaxSAT formulation of Ref.~\cite{Berent2024}, we use Z3~\cite{deMoura2008} to obtain global minimum-weight corrections. We compute the full logical success probability $P_{\mathrm{full}}$ and estimate the MW phase boundary by extrapolating the $P_{\mathrm{full}}=1/2$ contours in $1/L$. The resulting estimates exhibit a similar dependence on $r$ to those of the toric code. For both codes, the MW estimates already lie close to the principal-factor approximation to the ML threshold.

Recent work on weakly monitored quantum circuits with qubit loss has observed a  flow from percolation to Nishimori universality~\cite{Putz2025}. An interesting direction for future work is to investigate whether the optimal decoding in the present protocol exhibits analogous crossovers from geometric percolation in the toric code and homological percolation in the color code toward their respective critical behavior at the toric code infinite-randomness fixed point~\cite{Pandey2026} and its color code analogue.

\subsubsection*{Inference / Tunable learning}

The connection to logical inference also yields a bound on tunable-learning transitions~\cite{Eckstein2024,Eckstein2025}. 
Along the pure-$Z$ axis, the homological-percolation threshold provides a {\sl lower bound} on the weak-measurement learning threshold,
\begin{equation}
  (\gamma_Z)_{\mathrm c}\geq(r_Z)_{\mathrm c},
  \label{eq:weak_measurement_threshold}
\end{equation}
where $\gamma_Z\in[0,1]$ denotes the weak-measurement strength in the parametrization defined below. We write $\gamma=\gamma_Z$ throughout the following argument.

Intuitively, the weak-measurement record can be understood as coming from projective measurements performed on each qubit with probability $\gamma$, with random signs returned otherwise. In our projective protocol at $r=\gamma$, we additionally know which qubits were actually measured. This extra knowledge makes the logical sector easier to learn, which explains the threshold bound above.

\begin{figure}
    \centering
    \includegraphics[width=1.0\columnwidth]{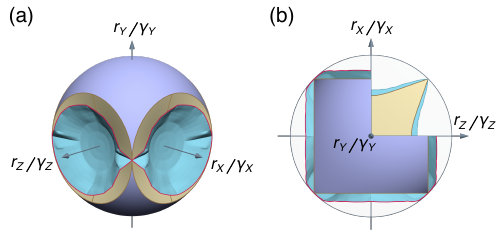}
    \caption{
    {\bf Overlay of toric-code phase diagrams.}
    Illustrated is a comparison of the spherical phase diagrams obtained for random-projective and weak-measurements,
    with the threshold boundaries depicted in yellow and cyan, respectively. 
    The two panels show (a) an oblique view, and (b) a cutaway viewed along the $Y$ axis.    
    What this comparison demonstrates is that the easy-to-calculate phase boundaries of the random-projective model provide
    guidance for the hard-to-calculate weak-measurement phase diagram:
    On a {\sl qualitative} level, the `percolation bites' in the random-measurement model get replaced by `Nishimori bites' in the 
    weak-measurement model allowing, for instance, to quickly predict the overall topology of the weak-measurement phase
    diagrams of all codes overviewed in Fig.~\ref{fig:family}. 
    On a {\sl quantitative} level, the percolation transitions of the random-measurement model put an lower bound on the Nishimori
    transitions of the weak-measurement model which is evident in both panels.}
    \label{fig:overlay}
\end{figure}

To see this picture more explicitly, consider a binary weak measurement of $Z_i$ with outcome $s_i=\pm1$. Its POVM effect can be decomposed as
\begin{equation}
  \begin{aligned}
    F^Z_{i,s_i}(\gamma)
    &=\frac{I+\gamma s_iZ_i}{2}\\
    &=\gamma P^Z_{i,s_i}+(1-\gamma)\frac{I}{2},
  \end{aligned}
\end{equation}
where $P^Z_{i,s_i}=(I+s_iZ_i)/2$ and $\sum_{s_i=\pm1}F^Z_{i,s_i}(\gamma)=I$. The first term gives a projective measurement outcome, while the second gives a random outcome that is equally likely to be $+1$ or $-1$, independent of the quantum state. For the complete record $s=(s_1,\ldots,s_n)$, this gives
\begin{equation}
  \begin{aligned}
    F^Z_s(\gamma)
    &=\bigotimes_{i=1}^nF^Z_{i,s_i}(\gamma)\\
    &=\sum_{M\subseteq \{1,\ldots,n\}}\gamma^{|M|}(1-\gamma)^{n-|M|}2^{-|U|}P^M_{s_M},
  \end{aligned}
  \label{eq:weak_measurement_effect}
\end{equation}
where $U=\{1,\ldots,n\}\setminus M$, and $P^M_{s_M}=\prod_{i\in M}P^Z_{i,s_i}$. The factor $2^{-|U|}$ accounts for the random signs reported on unmeasured qubits. Retaining $M$ gives the record of our random projective protocol at $r=\gamma$, together with irrelevant random signs on $U$. Discarding $M$ leaves precisely the weak-measurement outcome distribution.

Let $\Xi$ label a uniformly distributed complete logical-$Z$ sector, $M$ denote the random measurement pattern, and $S$ denote the random weak-measurement record. Revealing $M$ can only reduce the uncertainty about $\Xi$, while the random signs outside $M$ carry no additional information. Therefore,
\begin{equation}
H(\Xi\mid S)\geq H(\Xi\mid S,M)=H(\Xi\mid S_M, M),
\end{equation}
where $H(A|B)$ denotes the conditional Shannon entropy of $A$ given $B$. For these pure-$Z$ measurements, with all outcomes retained and no additional noise, the coherent information equals the remaining classical uncertainty about the logical sector (see Appendix~\ref{app:tunable_learning_thresholds}). Consequently,
\begin{equation}
  \begin{aligned}
    I_c^{\mathrm{weak}}(\gamma)
    &=H(\Xi\mid S)\\
    &\geq H(\Xi\mid S_M, M)
    =I_c^{\mathrm{proj}}(r=\gamma).
  \end{aligned}
\end{equation}
Assuming that both protocols exhibit a single sharp transition in coherent information for the same code family, this inequality implies the threshold bound in Eq.~\eqref{eq:weak_measurement_threshold}. The extension to general measurement directions is derived in Appendix~\ref{app:tunable_learning_thresholds}.

The pure-axis thresholds give an interpretation of the ``bites'' in the phase diagram---they determine how far they extend toward the origin along each axis. Our random-projective protocol gives lower bounds on the corresponding weak-measurement thresholds. For the toric code, the higher weak-measurement thresholds along the $X$ and $Z$ axes are reflected in the shallower bites in Fig.~\ref{fig:overlay}, while both protocols reach the maximal threshold along the $Y$ axis. These axial constraints help explain the overall relation between the two phase diagrams.

\section{Discussion}

One central result of this work is that logical-survival transitions in randomly monitored QEC codes are governed by homological connectivity, with ordinary geometric percolation appearing as a special realization. A measurement pattern collapses encoded information when products of the measured Pauli operators support a nontrivial logical Pauli class. In the pure-$Z$ toric code, stabilizer deformations coincide with local reroutings of an unbranched loop, and the problem reduces exactly to square-lattice bond percolation. The color code has the same transition point, $p_Z=1/2$, but its stabilizer relations permit branching and recombination beyond ordinary path connectivity. Its correlation-length exponent $\nu\simeq1.20$, together with the cluster exponents obtained from  large-system calculations, provides evidence for a universality class distinct from ordinary two-dimensional percolation. The smaller value of $\nu$ also implies a more rapid finite-size sharpening of the transition, favoring the survival of logical information near the threshold when other factors are comparable -- a potential practical advantage when implementing finite-distance codes on quantum processing platforms.

Beyond the code families discussed in Sec.~\ref{sec:wider_perspective} and illustrated in Fig.~\ref{fig:family}, the homological-percolation framework admits several natural extensions. More broadly, the problem is specified by combining a chain complex, which determines stabilizer equivalence and the nontrivial logical classes, with a random subspace generated by a chosen ensemble of commuting Pauli measurements and its sampling distribution~\footnote{Strictly speaking, a thermodynamic phase transition also requires specifying how the chain complexes and the generators of the random subspaces scale with system size.}. Closely related studies of random two-body $XX$ bond measurements in the color code, with the outcomes discarded, identified an emergent toric-code phase and formulated the transition in terms of percolation and homology on an associated triangular lattice~\cite{Kataoka2026,Kataoka2026-2}. Thus, different choices of local measurement generators can produce different percolation problems even for the same underlying code, opening a broader design space for measurement-induced transitions.

The Wen-plaquette example points to a related freedom within the same topological phase. Sublattice- and strip-dependent local Clifford rotations produce the XYZ and domain-wall color codes, respectively. Although these codes are locally equivalent to the conventional color code, their orientation relative to a physically fixed biased Pauli ensemble is different. The resulting reorganization of their logical supports substantially enhances code-capacity thresholds under biased Pauli noise, reaching $1/2$ in the infinite-bias limit~\cite{Miguel2023, Tiurev2024}. Applying the present measurement protocol to these variants would show how such local changes of Pauli frame reshape homological connectivity, phase boundaries, and critical behavior.

The construction also extends naturally to higher spatial dimensions, where the percolating logical supports may be membranes or higher-dimensional cycles, and to more general local complexes supporting branching or fractal logical structures~\cite{Duncan2023, Haah2011, Haah2013}. Qudit stabilizer codes provide another natural extension, replacing the binary Pauli algebra used here with its $q$-ary counterpart~\cite{Hostens2005,Ketkar2006}. As the qudit dimension increases, the number of inequivalent local Pauli measurement bases grows, which may enhance logical survival under uniform random measurements~\cite{Lee2026}. It remains to determine how the transition threshold and universality class depend on the qudit dimension.\\

Finally, let us again stress how knowledge about the percolation transitions in randomly monitored QEC codes can provide guidance to decoherence, monitoring, and inference settings and their respective transitions.
For this, we return to the comparison between the toric and color codes. Our principal statement that these two codes exhibit different universality classes for the random-monitoring transitions likely carries over to these other settings as well: Take, for instance, the optimal decoding transitions for these two codes when exposed to pure-$Z$ noise. While the thresholds of these transitions sit very close to one another~\cite{Katzgraber2009} and both map to an optimal Bayesian inference problem captured by Nishimori 
critical points, their universality classes are likely distinct since the two settings map to different statistical mechanics models: the random-bond versus the random three-body Ising model, respectively~\cite{Dennis2002,Katzgraber2009}, which exhibit different spin symmetries and domain-wall structures making distinct universality classes a natural possibility~\cite{Katzgraber2010}.

Likewise, minimum-weight decoding for these two codes probes the zero-temperature transitions of these underlying statistical models~\cite{Dennis2002,Thomas2011}. For the toric code, this transition is governed by a fixed point distinct from the Nishimori point and has recently been connected to infinite-randomness criticality~\cite{Pandey2026}. A distinct universality class for the color code is again a natural possibility. Indeed, a clear distinction appears in minimum-weight decoding complexity---under pure-$Z$ noise, exact minimum-weight decoding reduces to polynomial-time matching for the toric code, whereas it is NP-hard for the $(6,6,6)$ color code, with the hardness proof exploiting junctions of three differently colored strings~\cite{Dennis2002,Walters2026}. Although this worst-case complexity distinction per se does not determine the universality class, it highlights branching as a structural feature worth investigating.

The distinction between strings and branching string nets thus appears repeatedly in logical support, the domain walls of the associated statistical models, and decoding complexity. Our numerical evidence for distinct homological-percolation critical behavior gives further motivation to examine how this structure affects the corresponding decoding transitions. Exploring the connections and crossovers between homological percolation, decoding, and learning across different code families could clarify how the underlying code structure shapes this landscape of transitions.
\\

{\it Data availability.-- }
The numerical data shown in the figures are available on Zenodo~\cite{zenodo_repository}.

\acknowledgments
We thank Guo-Yi Zhu for discussions and collaboration on closely related projects. We acknowledge support, in part, by the Deutsche Forschungsgemeinschaft (DFG, German Research Foundation) under Germany's Excellence Strategy—Cluster of Excellence Matter and Light for Quantum Computing (ML4Q) EXC 2004/1 -- 390534769 as well as within the CRC network TR 183 (Project Grant No. 277101999) as part of subproject B01.
Our numerical simulations were performed on the Otus cluster at PC2 in Paderborn, the RAMSES cluster at RRZK Cologne, and the supercomputing systems at ISSP, the University of Tokyo.
D.S. acknowledges support from GP-Spin at Tohoku University.
B.H. acknowledges the hospitality of KITP at UC Santa Barbara, during the program ``Engineering Fractionalization in Quantum Many-Body Systems: From New Quantum Materials to Quantum Simulators'', where this work has been finalized.
This research was supported in part by grant NSF PHY-2309135 to the Kavli Institute for Theoretical Physics (KITP).
\\[-5mm]

\subsubsection*{Use of AI tools} 
ChatGPT and Codex (OpenAI; GPT-5.5, GPT-5.6, and GPT-6) were used to assist with literature searches, theoretical discussions, code development, and manuscript preparation. In particular, ChatGPT helped develop the homological-percolation formalism and the syndrome-space methods used to compute coherent information and homological-connectivity observables at large system sizes. This assistance notably included suggesting the use of binary-code coset labels to identify homological-connectivity classes. 
Writing assistance has been {\sl non-epistemic}, i.e., a non-generative, editorial use improving spelling, grammar, and phrasing.
High-level feedback (akin to a referee report), generated by Claude (Anthropic; Fable 5.1),
helped in the final polishing of the work.
The authors shaped and refined all scientific discussion, reviewed all AI-assisted code and arguments, and take full responsibility for the content of this work.

\appendix

\section{Homological percolation on chain and stabilizer complexes}
\label{app:homological_percolation}

Throughout this appendix, all vector spaces are over $\mathbb{F}_2$. The common object behind the examples in the main text is a chain complex
\begin{equation}
  C_2 \xrightarrow{\partial_2} C_1 \xrightarrow{\partial_1} C_0,
  \qquad
  \partial_1\partial_2=0 .
\end{equation}
The interpretation of the three vector spaces depends on the problem. In ordinary geometric percolation, $C_1$ is generated by edges and $C_2$ by plaquettes. In a CSS code, $C_1$ is generated by qubits and $C_2$ by stabilizer generators. In either case, the algebra is the same. Elements of $\ker\partial_1$ are closed objects, or cycles, while elements of $\operatorname{im}\partial_2$ are locally trivial cycles. The ambient first homology is
\begin{equation}
  H_1(C)=\frac{\ker\partial_1}{\operatorname{im}\partial_2} .
\end{equation}
Homological percolation asks whether a random subobject supports a nonzero class in this ambient homology.

For the purposes of this work, it is useful to formulate the random subobject as a random subspace
\begin{equation}
  W\subset C_1 .
\end{equation}
For ordinary bond percolation, $W$ is the span of the occupied edges. In a pure CSS measurement sector, such as single-qubit $Z$ measurements, $W$ is the span of the support vectors of the measured qubits. For mixed $X/Y/Z$ measurements, the same construction is applied in the full symplectic Pauli space, as described below.

The cycles supported by the random subobject are
\begin{equation}
  W\cap\ker\partial_1 .
\end{equation}
Their image in the ambient homology is
\begin{equation}
  \mathcal{L}(W)
  =
  \frac{(W\cap\ker\partial_1)+\operatorname{im}\partial_2}
       {\operatorname{im}\partial_2}
  \subset H_1(C) .
  \label{eq:abstract_hp_image}
\end{equation}
This image is the homological-percolation observable. Its rank,
\begin{equation}
  \rho(W)=\dim\mathcal{L}(W),
\end{equation}
counts how many independent ambient homology classes have representatives inside the random subobject. For a fixed class $[\ell]\in H_1(C)$ with $\ell\in\ker\partial_1$, the membership condition is
\begin{equation}
  [\ell]\in\mathcal{L}(W)
  \quad\Longleftrightarrow\quad
  \ell\in W+\operatorname{im}\partial_2 .
  \label{eq:abstract_membership}
\end{equation}
In words, the representative $\ell$ can be modified by locally trivial cycles until it lies inside the random subobject.

\subsection{Geometric homological percolation}

We first recover the familiar geometric example. Consider bond percolation on a square lattice with periodic boundary conditions. Let $C_1$ be generated by edges, $C_2$ by plaquettes, and $C_0$ by vertices. The map $\partial_2$ sends a plaquette to the sum of its boundary edges, while $\partial_1$ sends an edge to the sum of its two endpoint vertices. The condition
\begin{equation}
  \partial_1\partial_2=0
\end{equation}
is the algebraic statement that the boundary of a plaquette has no boundary: when one first takes the boundary edges of a plaquette and then takes the endpoints of those edges, every vertex appears twice and hence cancels over $\mathbb{F}_2$.

If $M$ is the set of occupied edges, then
\begin{equation}
  W=C_1(M)=\mathrm{span}_{\mathbb{F}_2}\{e:e\in M\}\subset C_1 .
\end{equation}
The space $W\cap\mathrm{ker}\,\partial_1$ consists of closed loops made only from occupied edges. Some of these loops are local and contractible. For example, if the four edges around a single plaquette are occupied, their sum is a closed loop, but it is also the boundary of that plaquette and hence belongs to $\operatorname{im}\,\partial_2$. Equation~\eqref{eq:abstract_hp_image} quotients out such local boundaries and keeps only the winding classes on the torus. Thus $\rho(W)>0$ means that the occupied edges contain a noncontractible loop.

More generally, let $T$ be a $d$-dimensional cell complex, where $d$ is the dimension of the ambient space, and suppose that the random process selects $i$-dimensional cells. Here $i$ is the dimension of the objects being percolated: for example, bonds have $i=1$, while plaquettes have $i=2$. The relevant local piece of the cellular chain complex is
\begin{equation}
  C_{i+1}(T) \xrightarrow{\partial_{i+1}} C_i(T) \xrightarrow{\partial_i} C_{i-1}(T),
\end{equation}
with $\partial_i\partial_{i+1}=0$. If $W\subset C_i(T)$ is the span of the selected $i$-cells, then $W\cap\mathrm{ker}\,\partial_i$ is the space of selected $i$-cycles. The homological-percolation observable is the image of these selected cycles in the ambient homology,
\begin{equation}
  \frac{(W\cap\mathrm{ker}\,\partial_i)+\mathrm{im}\,\partial_{i+1}}
       {\mathrm{im}\,\partial_{i+1}}
  \subset
  H_i(T)
  =
  \frac{\mathrm{ker}\,\partial_i}{\mathrm{im}\,\partial_{i+1}} .
\end{equation}

Equivalently, let $P$ be the random subcomplex obtained by taking the full $(i-1)$-skeleton of $T$ and adding the selected $i$-cells, and let $\phi:P\hookrightarrow T$ be the inclusion. Then the same observable is
\begin{equation}
  \mathrm{im}\left(\phi_*:H_i(P)\to H_i(T)\right),
\end{equation}
as in geometric homological percolation~\cite{Duncan2023}.

\begin{figure}
    \centering
    \vspace{0.5cm}
    \includegraphics[width=1.0\columnwidth]{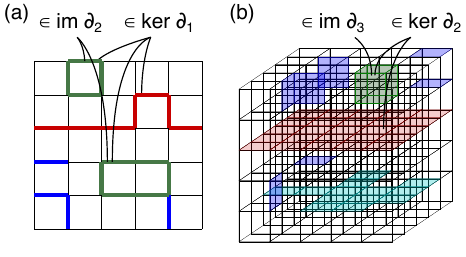}
    \caption{
    {\bf Schematics of geometric homological percolation.}
    The configurations are schematic examples, not samples drawn from a probability distribution.
    (a) For $(d,i)=(2,1)$, the selected objects are bonds on a square torus. Blue paths are not cycles, green loops are contractible cycles in $\mathrm{im}\,\partial_2\subset\mathrm{ker}\,\partial_1$, and the red loop lies in $\mathrm{ker}\,\partial_1$ but not in $\mathrm{im}\,\partial_2$, giving a nonzero class in $H_1(T^2)$.
    (b) For $(d,i)=(3,2)$, the selected objects are plaquettes on a cubic torus. Blue and cyan plaquette clusters are not cycles; the cyan cluster is drawn to span the system geometrically, but it still has a boundary. The green closed surface lies in $\mathrm{im}\,\partial_3\subset\mathrm{ker}\,\partial_2$, while the red closed surface lies in $\mathrm{ker}\,\partial_2$ but not in $\mathrm{im}\,\partial_3$, giving a nonzero class in $H_2(T^3)$.
    }
    \label{fig:geometric_homological_percolation}
\end{figure}

Figure~\ref{fig:geometric_homological_percolation} shows two examples of geometric homological percolation: $(d,i)=(2,1)$ in Fig.~\ref{fig:geometric_homological_percolation}(a) and $(d,i)=(3,2)$ in Fig.~\ref{fig:geometric_homological_percolation}(b). The case $(d,i)=(2,1)$ is the familiar bond-percolation problem on a two-dimensional torus, but viewed through homology. The blue bond configurations are open paths and therefore do not define homology classes. The green loops are cycles, but they are contractible because they bound plaquette regions; they are removed by quotienting by $\mathrm{im}\,\partial_2$. The red loop, by contrast, winds around the torus and survives as a nonzero element of $H_1(T^2)$.

The case $(d,i)=(3,2)$ is the corresponding plaquette-percolation problem on a three-dimensional torus. Here the selected objects are plaquettes, and the relevant cycles are closed surfaces. The blue plaquette patches have boundaries, and the cyan patch shows that even a cluster spanning the system geometrically is not enough if it is not closed. The green closed surface bounds a three-dimensional region and is therefore trivial in $H_2(T^3)$, whereas the red surface wraps nontrivially through the periodic system. Thus geometric homological percolation detects selected cycles that remain nonzero in the ambient homology, rather than merely large or spanning clusters.

\subsection{CSS homological percolation}

The CSS measurement problem is obtained by using the CSS chain complex in place of the cellular chain complex. We describe the $Z$ sector; the $X$ sector is obtained by interchanging $X$ and $Z$.

Let $C_1$ be the vector space generated by physical qubits, so an element of $C_1$ specifies the support of a $Z$-type Pauli operator. Let $C_2^Z$ be a label space for $Z$-stabilizer generators and $C_0^X$ a label space for $X$-stabilizer syndromes. The CSS complex is
\begin{equation}
  C_2^Z \xrightarrow{\partial_2^Z} C_1 \xrightarrow{\partial_1^Z} C_0^X .
  \label{eq:css_complex_app}
\end{equation}
Here $\mathrm{im}\partial_2^Z$ is the $Z$-stabilizer support space, while $\partial_1^Z$ maps a $Z$ support to its commutation syndrome with the $X$ stabilizers. Therefore $\ker\partial_1^Z$ is the space of $Z$ operators that commute with all $X$ stabilizers. The CSS commutation condition is exactly
\begin{equation}
  \partial_1^Z\partial_2^Z=0 .
\end{equation}
The $Z$-logical space is the homology
\begin{equation}
  H_1^Z=\frac{\ker\partial_1^Z}{\operatorname{im}\partial_2^Z} .
\end{equation}
This has the same form as closed loops modulo plaquette boundaries: normalizer $Z$ operators are the cycles, and $Z$ stabilizers are the locally trivial cycles.

Now consider single-qubit $Z$ measurements on a set of qubits $M$. The measurement record generates the $Z$-type support subspace
\begin{equation}
  W_Z=C_1(M)
  =
  \mathrm{span}_{\mathbb{F}_2}\{e_i:i\in M\}
  \subset C_1 ,
\end{equation}
where $e_i$ denotes the support vector of $Z_i$.
The measured $Z$-logical subspace is
\begin{equation}
  L_Z(W_Z)=
  \frac{(W_Z\cap\ker\partial_1^Z)+\operatorname{im}\,\partial_2^Z}
       {\operatorname{im}\,\partial_2^Z}
  \subset H_1^Z .
  \label{eq:css_hp_image_app}
\end{equation}
This is Eq.~\eqref{eq:abstract_hp_image} applied to the CSS complex. Its dimension counts the number of independent $Z$-logical eigenvalues fixed by the measurement outcomes.

For a fixed logical representative $\ell\in\ker\partial_1^Z$, the membership test becomes
\begin{equation}
  [\ell]\in L_Z(W_Z)
  \quad\Longleftrightarrow\quad
  \ell\in W_Z+\operatorname{im}\partial_2^Z .
  \label{eq:css_membership_app}
\end{equation}
Thus a logical $Z$ operator is measured if its representative can be written as a product of measured $Z$-type operators and $Z$ stabilizers.

For single-qubit measurements, Eq.~\eqref{eq:css_membership_app} has a useful cleaning form. Let $M$ be the measured qubits, let $U=\bar M$ be the unmeasured qubits, and let $R_U:C_1\to C_1(U)$ be restriction to $U$. Then $[\ell]\in L_Z(C_1(M))$ if and only if there exists a stabilizer pattern $a\in C_2^Z$ such that
\begin{equation}
  R_U(\ell+\partial_2^Z a)=0,
\end{equation}
or equivalently
\begin{equation}
  R_U\partial_2^Z a = R_U\ell .
  \label{eq:restricted_boundary_app}
\end{equation}
This says that the logical operator can be pushed away from every unmeasured qubit by multiplying stabilizers.

For the square-lattice toric code in the $Z$-only limit, $\partial_2^Z$ is the usual plaquette-boundary map. Therefore Eq.~\eqref{eq:restricted_boundary_app} reduces to the ordinary problem of whether the measured edges support a noncontractible loop. For the $(6,6,6)$ color code, $\partial_2^Z$ is instead the qubit--plaquette incidence matrix $D$, because each plaquette stabilizer acts on the qubits around a hexagon. Then Eq.~\eqref{eq:restricted_boundary_app} becomes
\begin{equation}
  D_Ua=\ell_U,
\end{equation}
where $D_U$ is obtained by restricting $D$ to unmeasured qubits. This is still homological percolation on the CSS complex, but it is not ordinary geometric connectivity percolation.

\subsection{Symplectic homological percolation for general Pauli measurements}

Mixed $X/Y/Z$ measurements do not stay within a single CSS sector. The corresponding chain complex is the stabilizer complex in the binary symplectic Pauli space. Modulo phases, an $n$-qubit Pauli operator is represented by
\begin{equation}
  v=(x|z)\in V=\mathbb{F}_2^{2n},
\end{equation}
where $x_i=1$ records an $X_i$ component and $z_i=1$ records a $Z_i$ component. The commutation relation is encoded by
\begin{equation}
  \omega\bigl((x|z),(x'|z')\bigr)=x\cdot z'+z\cdot x' .
\end{equation}
Let $S\subset V$ be the stabilizer space. Since stabilizers commute, $S$ is isotropic. Define the syndrome map
\begin{equation}
  \sigma:V\to S^\vee,
  \qquad
  \sigma(v)(s)=\omega(v,s) .
\end{equation}
The inclusion $\iota:S\hookrightarrow V$ and the syndrome map form a chain complex
\begin{equation}
  S \xrightarrow{\iota} V \xrightarrow{\sigma} S^\vee,
  \qquad
  \sigma\iota=0 .
  \label{eq:symplectic_complex_app}
\end{equation}
Its first homology is
\begin{equation}
  H_1^{\rm stab}=\frac{\ker\sigma}{\operatorname{im}\iota}
  =\frac{S^\perp}{S},
\end{equation}
which is precisely the logical Pauli space.

A commuting measurement realization generates an isotropic subspace
\begin{equation}
  W\subset V .
\end{equation}

The products of measured operators that commute with all stabilizers are the cycles
\begin{equation}
  W\cap\ker\sigma=W\cap S^\perp .
\end{equation}
Their image in the stabilizer homology is
\begin{equation}
  \mathcal L(W)=
  \frac{(W\cap S^\perp)+S}{S}
  \subset \frac{S^\perp}{S} .
  \label{eq:symplectic_image_app}
\end{equation}
This is Eq.~\eqref{eq:abstract_hp_image} applied to the stabilizer complex~\eqref{eq:symplectic_complex_app}. The rank
\begin{equation}
  \rho(W)=\dim\mathcal L(W)
\end{equation}
is the number of independent logical Pauli eigenvalues fixed by the measurement record. Because the measured operators commute, $\mathcal L(W)$ is isotropic in the logical symplectic space, so $0\le \rho(W)\le k$ for a code encoding $k$ logical qubits. Complete logical collapse corresponds to $\rho(W)=k$.

For a target logical class $[\ell]\in S^\perp/S$, the membership criterion is
\begin{equation}
  [\ell]\in\mathcal L(W)
  \quad\Longleftrightarrow\quad
  \ell\in W+S .
  \label{eq:symplectic_membership_app}
\end{equation}
Thus a logical Pauli operator is measured exactly when some representative of that logical class can be expressed as a product of measured Pauli operators and stabilizers. For pure $Z$-type or pure $X$-type CSS measurements, Eq.~\eqref{eq:symplectic_image_app} reduces to the corresponding CSS formula, such as Eq.~\eqref{eq:css_hp_image_app}. For mixed $X/Y/Z$ measurements, it keeps track of the $X$ and $Z$ components simultaneously.

\section{Finite-size scaling along the phase boundaries}
\label{app:phase_boundary_scaling}

\begin{figure}
    \centering
    \includegraphics[width=1.0\columnwidth]{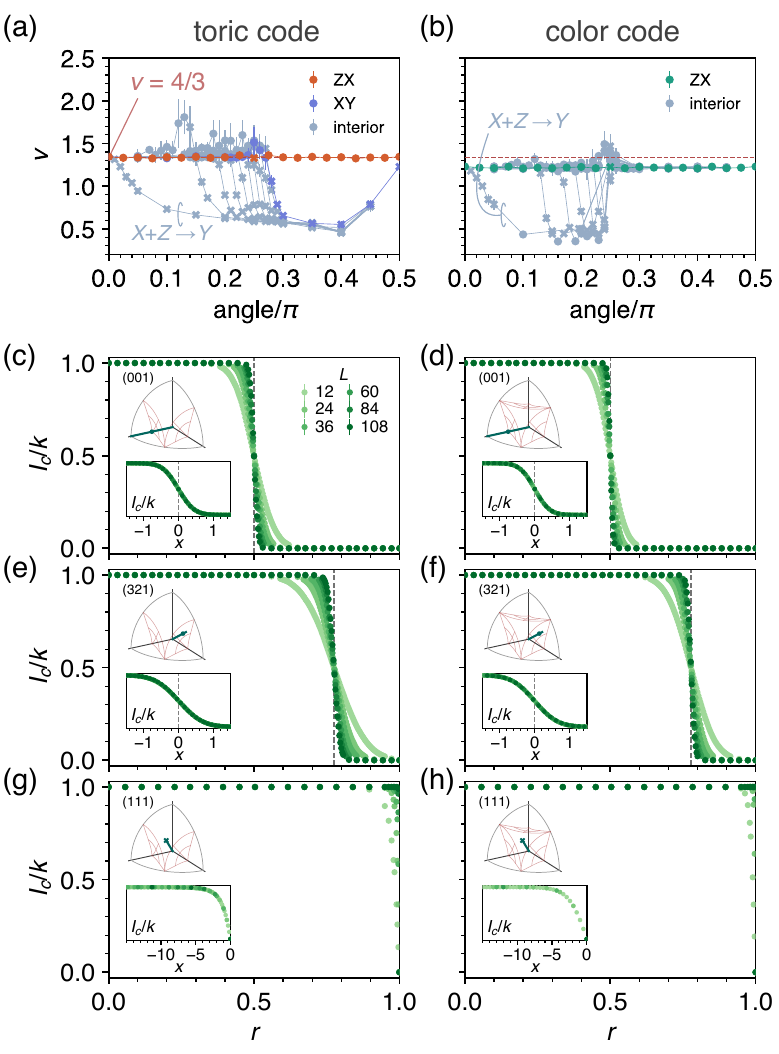}
    \caption{
    {\bf Critical scaling along selected phase-boundary directions.}
    (a,b) Correlation-length exponent $\nu$ extracted from finite-size scaling of the normalized coherent information for (a) the toric code and (b) the color code. Light blue denotes interior cuts at fixed azimuth. The curve labeled $X+Z\to Y$ follows the $r_X=r_Z$ cut from the $(101)$ direction toward the pure-$Y$ direction $(010)$ as the plotted angle increases. Cross symbols correspond to the phase-boundary fits compatible with $r_{\mathrm c}=1$.
    (c--h) Normalized coherent information $I_c/k$, where $k$ is the number of encoded logical qubits, versus total measurement probability $r$ for the toric code (left) and color code (right). The rows correspond to the $(001)$, $(321)$, and $(111)$ rays, respectively. Insets show the corresponding finite-size collapses versus $x=(r-r_{\mathrm c})L^{1/\nu}$.
    }
    \label{fig:coherent_information}
\end{figure}

Figure~\ref{fig:coherent_information} presents the scaling data and representative collapses underlying the compact summary in Fig.~\ref{fig:numerical_data}(d). Away from directions where two collapse branches approach one another, the toric-code estimates remain close to the ordinary two-dimensional percolation value $\nu=4/3$, whereas the color-code estimates remain close to $\nu\simeq1.2$. The distinction found on the pure axes therefore persists along representative mixed-measurement cuts.

Near the meeting of two collapse branches, a radial scan intersects the phase boundary at a shallow angle. Varying $r$ then produces only a small displacement normal to the boundary, making $r_{\mathrm c}$ and $\nu$ strongly correlated in a finite-size fit. This produces the pronounced peaks, enlarged uncertainties, and stronger size dependence observed near the balanced directions.

Directions for which the fitted $r_{\mathrm c}$ is compatible with unity exhibit a different endpoint scaling. In the toric-code pure-$Y$ limit, the crossover is controlled by $\mathcal O(L)$ translated $Y$-only winding strings of weight $2L$, while products of the corresponding line operators also yield dense $Y$-only representatives with support proportional to $L^2$~\cite{Tuckett2019, Botzung2025, Lee2026}. The total occurrence rate of the minimal winding strings is proportional to $Lp_Y^{2L}$, giving $1-p_Y\sim(\log L)/L$ along a fixed-survival contour and a power-law fit that drifts toward $\nu=1$. For directions close to but distinct from the pure-$Y$ ray, the effective exponent instead approaches $\nu\simeq1/2$. This is consistent with a crossover controlled by the number of unmeasured qubits on such an extensive support, for which the natural scaling variable is $(1-r)L^2$. The angular dependence thus suggests a crossover from rigid winding-string events on the pure-$Y$ ray to sharper, area-like behavior nearby.

\section{Floquet code numerical results}
\label{app:Floquet}

\begin{figure}[h!]
    \centering
    \includegraphics[width=1.0\columnwidth]{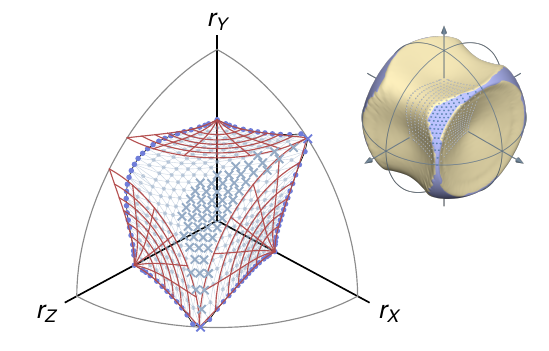}
    \caption{
    {\bf Numerical phase diagram of the Floquet code.}
    Red wireframes show the semi-analytic construction from the pure-axis thresholds, while colored symbols show numerical boundary estimates obtained from finite-size scaling of the coherent information. Crosses mark estimates compatible with $r_{\mathrm c}=1$ within uncertainty. The inset shows a 3D visualization of the phase diagram.
    }
    \label{fig:floquet_numerics}
\end{figure}

In Fig.~\ref{fig:floquet_numerics} we show a detailed numerical phase diagram of the Floquet code, complementing the semi-analytical phase diagram
shown in Fig.~\ref{fig:family}(i) of the main text.
Note in particular that the numerical simulations indicate a logical collapse occurring at $r_{\mathrm c}\approx0.83$ in the direction $r_X=0$ and $r_Y=r_Z$, below the semi-analytic prediction $r_{\mathrm c}=1$. In the hexagonal toric-code description of the post-$XX$ state, $X$ strings transport $e$ excitations, while strings built from $YY$ or $ZZ$ operators on the fixed bonds transport $m$ excitations~\cite{Hastings2021}. Since these two bond operators differ only by the fixed $XX$ check, up to a known sign, $Y$ and $Z$ measurements can contribute to the same magnetic logical string. Their combined effect is therefore not captured by the pure-axis thresholds alone.

\section{Tunable-learning thresholds}
\label{app:tunable_learning_thresholds}

We extend the comparison between weak and random projective measurements to general measurement directions. Let $\boldsymbol{\gamma}=(\gamma_X,\gamma_Y,\gamma_Z)$, with $\gamma_\alpha\geq0$ and $\gamma_X+\gamma_Y+\gamma_Z\leq1$. The weak-measurement POVM effects admit the decomposition
\begin{equation}
\begin{aligned}
F_{i,s_i}^{\mathrm{weak}}(\boldsymbol{\gamma})
&=\frac{I+s_i(\gamma_X X_i+\gamma_Y Y_i+\gamma_Z Z_i)}{2}\\
&=\left(1-\sum_{\alpha=X,Y,Z}\gamma_\alpha\right)\frac{I}{2}
+\sum_{\alpha=X,Y,Z}\gamma_\alpha P_{i,s_i}^{\alpha},
\end{aligned}
\label{eq:general_weak_measurement_effect}
\end{equation}
where $P_{i,s_i}^{\alpha}=(I+s_i\sigma_i^\alpha)/2$. Thus, the weak-measurement record can be understood as coming from projective $X_i$, $Y_i$, or $Z_i$ measurements performed with probabilities $\gamma_X$, $\gamma_Y$, and $\gamma_Z$, respectively, with independent fair signs returned otherwise. Revealing which alternative occurred on each qubit gives our random projective protocol with $p_\alpha=\gamma_\alpha$, together with irrelevant random signs on unmeasured qubits. Discarding the measurement choices recovers the weak-measurement outcome distribution.

Let $b$ denote the complete projective record, including measurement locations, Pauli axes, and outcomes. Let $T(s\mid b)$ be the classical channel that produces the sign string $s$ by adding independent fair signs on unmeasured qubits and discarding the measurement choices. The key observation is that, after tracing out the physical code, the conditional reference state depends only on the POVM effect, independently of how the measurement is implemented. For the initial state $|\Phi\rangle_{RQ}$, its unnormalized form is
\begin{equation}
\tau_R(x)
=\operatorname{Tr}_Q\!\left[
(I_R\otimes F_x)|\Phi\rangle\langle\Phi|
\right].
\label{eq:reference_state_from_effect}
\end{equation}
This expression is linear in $F_x$. The POVM decomposition therefore directly implies
\begin{equation}
\tau_R^{\mathrm{weak}}(s)
=\sum_b T(s\mid b)\tau_R^{\mathrm{proj}}(b).
\label{eq:unnormalized_reference_comparison}
\end{equation}
Here the unnormalized states include the probabilities of their records, including the random measurement choices in the projective protocol. Writing $p_b=\operatorname{Tr}\tau_R^{\mathrm{proj}}(b)$ and $p_s=\operatorname{Tr}\tau_R^{\mathrm{weak}}(s)$, normalization gives
\begin{equation}
\begin{aligned}
\rho_R^{\mathrm{weak}}(s)
&=\sum_b p(b\mid s)\rho_R^{\mathrm{proj}}(b),\\
p(b\mid s)&=\frac{T(s\mid b)p_b}{p_s}.
\end{aligned}
\label{eq:conditional_reference_comparison}
\end{equation}
Thus, each conditional reference state of the weak protocol is a mixture of the reference states associated with the more informative projective record.

The entropy of this mixture is at least the average entropy of its components. Averaging over the weak-measurement records and using Eq.~\eqref{eq:coherent_information_record}, we obtain
\begin{equation}
\begin{aligned}
I_c^{\mathrm{weak}}(\boldsymbol{\gamma})
&=\sum_s p_s S\!\left(\rho_R^{\mathrm{weak}}(s)\right)\\
&\geq\sum_{s,b}p_s p(b\mid s)
S\!\left(\rho_R^{\mathrm{proj}}(b)\right)\\
&=\sum_b p_b S\!\left(\rho_R^{\mathrm{proj}}(b)\right)\\
&=I_c^{\mathrm{proj}}(\boldsymbol p=\boldsymbol{\gamma}).
\end{aligned}
\label{eq:general_coherent_information_bound}
\end{equation}

In the pure-$Z$ limit, the conditional reference states are diagonal in the logical-$Z$ basis, so Eq.~\eqref{eq:coherent_information_record} reduces to $I_c=H(\Xi\mid C)$, recovering the classical formulation used in the main text.

To compare thresholds, fix a unit direction $\hat{\boldsymbol n}$ in the nonnegative octant and write $\boldsymbol{\gamma}=\gamma\hat{\boldsymbol n}$, so that $\gamma=\|\boldsymbol{\gamma}\|_2$. Let $\boldsymbol p_{\mathrm c}(\hat{\boldsymbol n})$ denote the boundary of the projective protocol's logical-preservation region along the same ray. Assuming a single sharp transition in both protocols, with the same coherent-information criterion, we obtain
\begin{equation}
\gamma_{\mathrm c}(\hat{\boldsymbol n})
\geq\|\boldsymbol p_{\mathrm c}(\hat{\boldsymbol n})\|_2.
\label{eq:general_weak_measurement_threshold}
\end{equation}
The weak-measurement POVM is well defined throughout $\gamma_X^2+\gamma_Y^2+\gamma_Z^2\leq1$, whereas the decomposition above requires $\gamma_\alpha\geq0$ and $\gamma_X+\gamma_Y+\gamma_Z\leq1$. The weak-measurement transition itself may lie outside this comparison region.

An alternative comparison is obtained by projectively measuring the same observable $O_{\theta,\phi}$ as the weak protocol, with probability $r$. The same record decomposition then applies at $r=\gamma$ throughout $0\leq\gamma\leq1$, giving $\gamma_{\mathrm c}(\theta,\phi)\geq r_{\mathrm c}^{O}(\theta,\phi)$ under the same sharp-transition assumption. For generic directions, this involves a different projective protocol from the random-Pauli measurements studied here.

\bibliography{ref}

\end{document}